\documentclass[aps,prl,preprint,amssymb,showpacs,onecolumn,superscriptaddress,notitlepage]{revtex4-1}
\usepackage{bm}
\usepackage{graphicx}
\usepackage{dcolumn}
\usepackage{braket}
\usepackage{natbib}
\usepackage{amsmath}
\usepackage{color}
\usepackage{ulem}
\usepackage{siunitx}
\usepackage{caption, setspace}
\graphicspath{{figs/}}
\usepackage[colorlinks=true,citecolor=blue,urlcolor=blue,linkcolor=blue,bookmarksopen]{hyperref}

\usepackage{soul}

\usepackage{natbib}
\usepackage{amsmath}
\usepackage{color}
\usepackage{ulem}
\usepackage{siunitx}
\graphicspath{{figs/}}

\begin{document}
\title{Polariton Bose-Einstein condensate from a Bound State in the Continuum}
\author{V. Ardizzone}
\affiliation{CNR Nanotec, Institute of Nanotechnology, via Monteroni, 73100, Lecce}
\author{F. Riminucci}
\affiliation{Molecular Foundry, Lawrence Berkeley National Laboratory, One Cyclotron Road, Berkeley, California, 94720, USA}
\affiliation{Dipartimento di Fisica, Universit\`a del Salento, Strada Provinciale Lecce-Monteroni, Campus Ecotekne, Lecce 73100, Italy}
\author{S. Zanotti}
\affiliation{Dipartimento di Fisica, Universit\`a di Pavia, via Bassi 6, 27100 Pavia (IT)}
\author{A. Gianfrate}
\affiliation{CNR Nanotec, Institute of Nanotechnology, via Monteroni, 73100, Lecce}
\author{M. Efthymiou-Tsironi}
\affiliation{CNR Nanotec, Institute of Nanotechnology, via Monteroni, 73100, Lecce}
\author{D. G. Suarez-Forero}
\affiliation{CNR Nanotec, Institute of Nanotechnology, via Monteroni, 73100, Lecce}
\affiliation{Dipartimento di Fisica, Universit\`a del Salento, Strada Provinciale Lecce-Monteroni, Campus Ecotekne, Lecce 73100, Italy}
\author{F. Todisco}
\affiliation{CNR Nanotec, Institute of Nanotechnology, via Monteroni, 73100, Lecce}
\author{M. De Giorgi}
\affiliation{CNR Nanotec, Institute of Nanotechnology, via Monteroni, 73100, Lecce}
\author{D. Trypogeorgos}
\affiliation{CNR Nanotec, Institute of Nanotechnology, via Monteroni, 73100, Lecce}
\author{G. Gigli}
\affiliation{CNR Nanotec, Institute of Nanotechnology, via Monteroni, 73100, Lecce}
\affiliation{Dipartimento di Fisica, Universit\`a del Salento, Strada Provinciale Lecce-Monteroni, Campus Ecotekne, Lecce 73100, Italy}
\author{K. Baldwin}
\affiliation{PRISM, Princeton Institute for the Science and Technology of Materials, Princeton University, Princeton, New Jersey 08540, USA}
\author{L. Pfeiffer}
\affiliation{PRISM, Princeton Institute for the Science and Technology of Materials, Princeton University, Princeton, New Jersey 08540, USA}
\author{D. Ballarini}
\affiliation{CNR Nanotec, Institute of Nanotechnology, via Monteroni, 73100, Lecce}
\author{H. S. Nguyen}
\affiliation{Univ Lyon, ECL, INSA Lyon, CNRS, UCBL, CPE Lyon, INL, UMR5270, 69130 Ecully, France} 
\affiliation{Institut Universitaire de France (IUF)}
\author{D. Gerace}
\email{dario.gerace@unipv.it}
\affiliation{Dipartimento di Fisica, Universit\`a di Pavia, via Bassi 6, 27100 Pavia (IT)}
\author{D. Sanvitto}
\email{daniele.sanvitto@nanotec.cnr.it}
\affiliation{CNR Nanotec, Institute of Nanotechnology, via Monteroni, 73100, Lecce}

\maketitle

\textbf{
{Optical bound states in the continuum (BIC) are peculiar topological states that, when realized in a planar photonic crystal lattice, are symmetry-protected from radiating in the far field despite lying within the light cone, i.e., in the energy-momentum dispersion region for which radiation can propagate out of the lattice plane. These BICs possess an invariant topological charge given by the winding number of the polarization vectors, similarly to vortices in quantum fluids, such as superfluid helium and atomic Bose-Einstein condensates. In spite of several reports of optical BICs in patterned dielectric slabs with evidence of lasing, their potential as topologically protected states with theoretically infinite lifetime has not been fully exploited, yet. Here
we show Bose-Einstein condensation of polaritons, hybrid light-matter excitations, occuring in a BIC thanks to its peculiar non-radiative nature.
The combination of the ultra-long BIC lifetime and the tight confinement of the waveguide geometry allow to achieve an extremely low threshold density for condensation, which is not reached in the dispersion minimum but at a saddle point in reciprocal space. By bridging  bosonic condensation and symmetry-protected radiation eigenmodes, we unveil new ways of imparting topological properties onto macroscopic quantum states with unexplored dispersion features. Such an observation may open a route towards energy-efficient polariton condensation in cost-effective integrated devices, ultimately suited for the development of hybrid light-matter optical circuits.}
} 

{Optical BICs,\cite{Hsu2016} although lying in the continuum radiation spectrum for which it is unavoidable to leak into free space, are in fact characterized by an infinite lifetime  and a quantized topological charge.\cite{Zhen2014} While BICs were predicted back in 1929 as a mathematical solution of the Schrodinger equation with \textit{ad hoc} potentials,\cite{Neumann1993} they have later been related to destructive interference,\cite{Friedrich1985} and observed in different scenarios ranging from electronic transport\cite{Capasso1992} to acoustic, fluids, and electromagnetic waves.\cite{Hsu2016} More recently, there has been a fervid experimental activity to realise optical BICs in several photonic platforms\cite{MermetLyaudoz2019} in which room temperature lasing could also be achieved\cite{Kodigala2017}. It is therefore natural to wonder if such a topologically protected state, with a supposedly infinite lifetime, can serve as ground state for the realisation of Bose-Einstein condensation (BEC) for collective solid state excitations. The latter was first observed in a semiconductor microcavity with highly reflecting mirrors \cite{Kavokin2008}, where strong light-matter coupling gives rise to hybrid elementary excitations of excitons and photons, named \textit{polaritons}, whose lifetime may range from a few to at most a hundred picoseconds. To reach such relatively long lifetimes, and sufficiently high electron densities, the use of many quantum wells (QWs) and several layers of distributed Bragg reflectors have to be used. Quantum coherent phenomena arise once pumping has provided a certain critical density and the threshold for BEC is reached. In the past decades the literature has flourished with the observation of several such effects, from typical BEC,\cite{Kasprzak2006,Balili2007} even at room temperature,\cite{Dusel2020} to superfluidity,\cite{Amo2009,Lerario2017} quantized vortices,\cite{Lagoudakis2008,Lagoudakis2009,Sanvitto2010} soliton
propagation,\cite{Amo2011,Walker2017} as well as topology-related helicity.\cite{LiuScience2020} However, most of these achievements are based on microcavity samples that require significant fabrication efforts\cite{Amo2011}. In fact, a recently explored polaritonic platform makes use of planar waveguides, which allow for tighter vertical confinement and very long lifetimes of propagating polaritons, since they rely on total internal reflection.\cite{Walker2013} Moreover, it has been recently demonstrated that nonlinearities can be easily tuned in polariton waveguides by using external electric fields to allow the formation of dipolar excitons with long range Coulomb interactions.\cite{Rosenberg2016,Rosenberg2018} Such interactions can be almost one order of magnitude larger than those driven by exchange coupling,\cite{Togan2018,SuarezForero2020} with strong implication for  polaritons in quantum information processing. In addition, such heterostructures are much simpler to grow and easier to integrate in, e.g., hybrid solutions with silicon electronics. However, the propagating photonic mode undergoing strong light-matter coupling with the excitonic resonance in a waveguide has no energy minimum in the dispersion, which does not allow, in principle, to achieve polariton condensation by accumulation of particles.

Here we demonstrate that strong exciton-radiation coupling to a BIC allows to reach polariton BEC 
by just using a patterned  waveguide. In order to realise a polariton BEC in a BIC, we use a one-dimensional (1D) sub-wavelength grating etched on top of a multi-QW GaAs heterostructure, thus folding the propagating polariton guided modes close to zero in-plane momentum with no in-plane group velocity. Hence, in spite of observing coherent emission from a waveguide due to the presence of the 1D grating, losses are almost suppressed due to the coupling to a photonic BIC of ideally infinite Q-factor. 
We show that polariton condensation is reached at this critical point with a minimal excitation density. 
Notice that in a similar structure, but in the absence of BIC, polariton condensation does not occur before regular photon lasing is established.\cite{Turnbull2001} Moreover, such a condensed state does not appear in the minimum but rather at a saddle point of the planar dispersion (corresponding to the polariton BIC condition). In our samples the polariton BEC is reached with a threshold energy density per pulse in the order of a few $\mu$J/cm$^{2}$, which is considerably lower than both conventional photon lasers - even those recently obtained at a BIC\cite{Kodigala2017} - and polariton condensates in microcavities.\cite{Sanvitto2016} 
In addition, we directly measure the polarization vortex pattern associated to the BIC topological charge,\cite{Doeleman2018} showing that BEC in BICs is also a viable route for transferring fundamental topological properties to polariton condensates.
}

\begin{figure}[htbp]
\captionsetup{font={small,stretch=0.6},justification=justified}
    \centering
    \includegraphics[height=0.55\textheight, keepaspectratio]{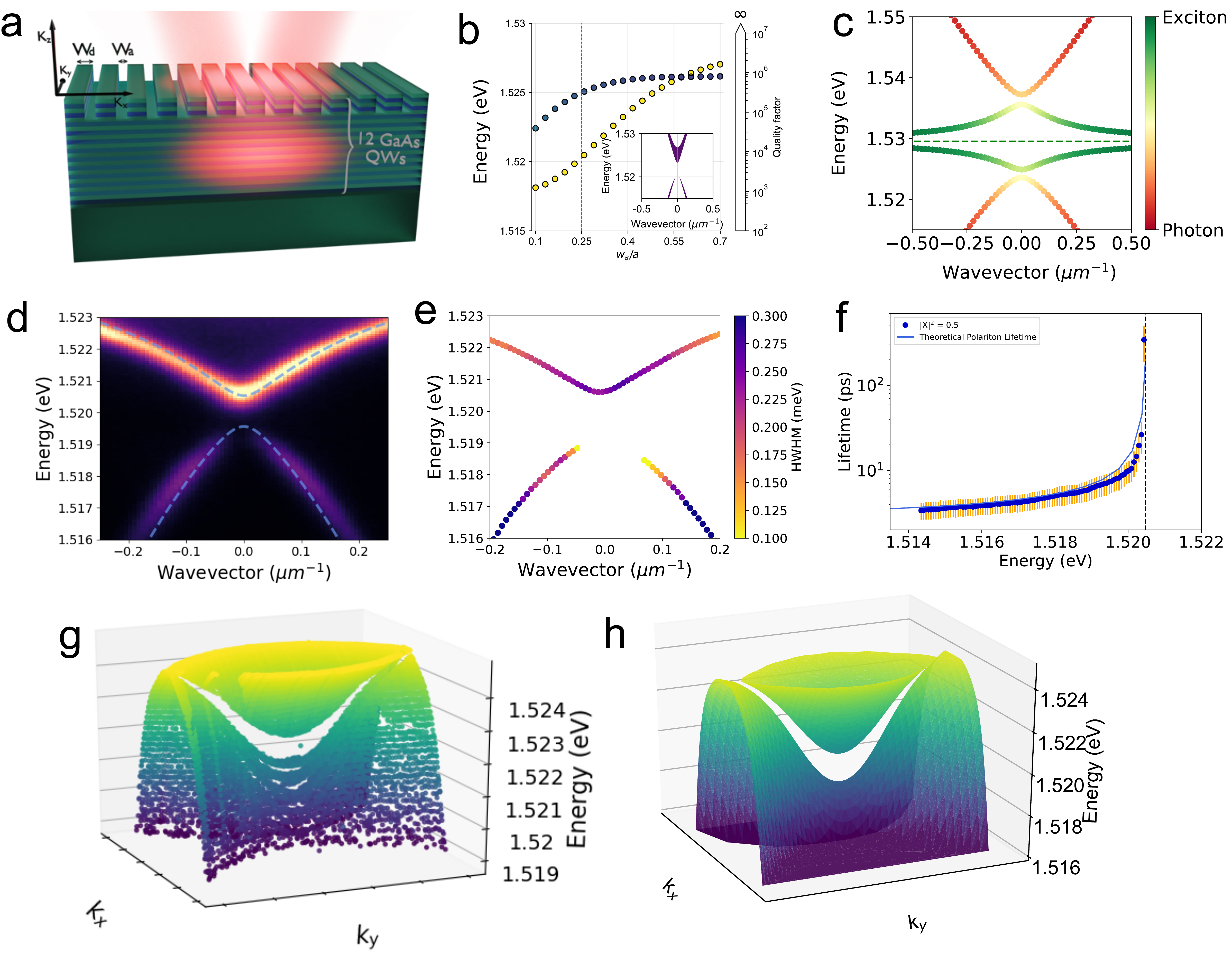}
\caption{\textbf{Polariton BIC.} 
\textbf{a} Artistic representation of the polariton waveguide with partially etched 1D lattice, and corresponding reference frame.
\textbf{b} Dependence of the upper and lower band extrema in $k_x=0$ on the grating air fraction ($w_a/a$), with a color code corresponding to the calculated Q-factor, as calculated by FDTD (see Methods); The inset shows the calculated dispersion of the grating modes (assuming no exciton resonance), the line thickness represents the width of the corresponding purely photonic resonance for $w_a=0.25a$ (highlighted with a vertical line in the plot).
\textbf{c} Polariton dispersion as a function of wave vector along the propagation direction in the relevant energy range around the excitonic transition (green dashed line), calculated from a coupled oscillators model, taking the FDTD results for the dispersion as the photonic components and then coupling them to the excitonic resonance; 
the colors represent the excitonic fraction for each mode. \textbf{d} Angular resolved PL emission under non-resonant excitation from a grating having a pitch $a\sim 240$ nm and fill factor $FF\sim 0.7$. The dark spot at $E \sim 1.519$ eV on the lower polariton branch comes from the polariton BIC.  A model (blue dashed line) of two coupled resonances hybridized with the exciton (as in panel c) is used to fit the polariton dispersion (i.e., the blue dashed line corresponds to the lower polariton branches of panel c); \textbf{(e)} Experimentally extracted peak energies, and corresponding HWHM (color scale) from the two polariton modes visible in (d) as a function of $k_x$; The points closest to $k_x\sim 0$ cannot be characterized due to the lack of signal from the dark state.  \textbf{(f)} Energy resolved lifetime of propagating polaritons from the branch hosting the BIC mode. 
\textbf{(g)} {Dispersion of the polariton modes as a function of $k_{x}$ and $k_{y}$, extracted from experimental spectra}. The dispersion of the lower branch clearly forms a saddle, with a minimum along $k_{y}$ and a maximum along $k_x$. The color code  represents energy increasing from dark to light. \textbf{(h)} Calculated polariton dispersion along $k_{x}$ and $k_{y}$, obtained by {the coupled oscillators model}, as in panels (c) and (d). }
\label{sketch}
\end{figure}

{Figure~\ref{sketch} presents an overview of the light emitting structure considered in this work, and a full characterization of its linear optical properties. {Further theoretical and experimental details are reported in the Supplementary Material (SM) file.} In particular, we hereby consider a multi-layer planar waveguide consisting of 12 GaAs QWs in AlGaAs barriers partially etched with a 1D grating for the first 5 layers, as represented in Fig.~\ref{sketch}a. The 1D grating is characterized by fill factor $FF=w_d/a$ ($a$ is the grating pitch) and air fraction $w_a/a=1-FF$. The grating FF can be used to tune the symmetry properties of photonic resonances, and engineer their intrinsic losses.\cite{GeracePRE2004}
Two such resonances (or photonic modes) are plotted in the inset of Fig.~\ref{sketch}b as a function of the wave vector along the propagation direction of the grating, $k_x$, and corresponding to $FF=0.75$. Line thickness in the plot is proportional to the calculated spectral width (i.e., inverse quality factor) of each resonance, showing that the lowest energy mode has zero intrinsic linewidth at $k_x=0$ , i.e., it is a symmetry-protected BIC, while the branch with a minimum {at} $k_x=0$ is lossy and displays a low Q-factor. The calculated electric field profiles show that the BIC mode is actually antisymmetric for mirror reflection against the center of the elementary cell, which yields suppression of the vertical losses by destructive interference,\cite{Hsu2016} while the bright mode is symmetric. 
This property depends on the grating air fraction, as shown in Fig.~\ref{sketch}b, in which $Q>10^7$ corresponds to the resonance loss becoming negligibly small while approaching the BIC in the simulation, thus unequivocally identifying this  mode.}\\  
Strong light-matter coupling between a QW excitonic transition and such photonic modes leads to the formation of hybrid excitations defined \textit{BIC polaritons}.\cite{KoshelevPRB2018,Kravtsov2020,Lu2020} The coupling of these polaritons to the radiative continuum is strictly forbidden due to the symmetry mismatch inherited from the photonic BIC. Thus, they possess infinite radiative Q-factor and are perfectly localized within the few hundreds nm-thick grating. Such perfect confinement is a new paradigm compared to traditional exciton-polaritons in planar microcavity samples with thick distributed Bragg reflectors. On the other hand, the hybrid nature of the BIC polaritons implies that they may exhibit non-radiative losses inherited from the excitonic resonances, and their lifetime is inevitably bounded by these losses, unavoidable in semiconductor systems yet very low in GaAs-based QWs. 
In our sample, both photonic modes of Fig.~\ref{sketch}b strongly couple to the excitonic transition. The coupling and the resulting eigenmodes can be described by a model of 4 coupled oscillators. The calculated polariton dispersions are shown in Fig. \ref{sketch}c. 
{Angular and energy resolved photoluminescence (PL) is reported in Fig.~\ref{sketch}d showing a direct measurement of the polariton dispersion at energies below the exciton resonance}. Around $E \sim 1.520$ eV an avoided crossing between the two counter-propagating polariton modes is clearly visible, corresponding to the purely photonic gap induced by the periodic dielectric modulation (see, e.g., the inset of Fig. \ref{sketch}b).
Here, it is relevant to stress that the {two polariton modes appearing with a maximum and a minimum at $k_x=0$, respectively, possess very different properties:}
while the former mode is clearly dark, the latter remains bright at normal incidence. The blue dashed line of Fig.~\ref{sketch}d represents the real part of the eigenvalues calculated by diagonalizing the corresponding $4\times 4$ Hamiltonian. In Fig.~\ref{sketch}e we also plot the energy of the polariton modes extracted from  Fig.~\ref{sketch}d, with a color scale representing the corresponding half width at half maximum (HWHM), after having taken into account the spectral resolution of the imaging system. The narrowing of the HWHM around $k_x\sim 0$ clearly shows the formation of a polariton BIC, arising from the strong exciton-photon coupling with the purely photonic BIC mode theoretically predicted in Fig.~\ref{sketch}b. Notice that while approaching $k_x = 0$, the linewidth becomes smaller than the spectral resolution and cannot be reliably extracted below a certain angle.\\ 
While the linewidth narrowing is a clear evidence of a polariton BIC, the measured linewidth values are not directly related to polariton lifetimes due to either the limited spectrometer resolution or the unavoidable exciton inhomogeneous broadening, {particularly evident} in our sample containing 12 QWs. {Therefore, in order to extract the {actual} polariton lifetimes from such a dark state we have used a different technique, taking advantage of the control on the dispersion relation and polariton propagation along the patterned waveguide. With time of flight measurements we were then able to establish the asymptotic lifetime of the dark BIC mode.} Lifetimes can then be determined  for each wavevector from the polariton propagation lengths, knowing the corresponding polariton group velocity extracted from the experimental {dispersion shown} in Fig.~\ref{sketch}d. 
Polariton lifetimes measured with this technique are reported in Fig.~\ref{sketch}f, showing that {when the energy of the propagating mode approaches} the polariton BIC, the measured lifetime rapidly increases reaching values exceeding $\sim300$ ps, yet remaining radiatively dark. Despite the asymptotic lifetime could even suggest an infinite value, the polariton BIC has a finite lifetime due to the hybridisation with excitons and their intrinsic non-radiative losses. {The latter, however, are known to be relatively small in GaAs QWs}. We stress once more that due to the dark nature of the polariton BIC we cannot measure lifetimes at the exact polariton BIC energy, and the measurements reported here should be considered as a lower bound for polariton BIC lifetimes. \\
Remaining in the linear regime, it is interesting to study the full polariton dispersion in reciprocal space, around the BIC condition, which is conveniently done by resolving the PL emission along both $k_{x}$ and $k_{y}$. The resulting experimental dispersion is  visualized in Fig.~\ref{sketch}g with a 3D plot, in which the two polariton branches are both represented. This plot clearly shows that the lower branch {displays a saddle-like dispersion}, with the polariton BIC sitting on a maximum along $k_{x}$ while being on a minimum along $k_{y}$. This peculiar configuration is well reproduced by the model of the coupled counter-propagating modes, shown in Fig.~\ref{sketch}h. These are obtained again by coupling the {calculated} photonic modes to the excitonic transition, and then diagonalizing the full Hamiltonian in both reciprocal space directions. {We also notice that due to its saddle-like shape the lower polariton branch gets closer and closer to the excitonic reservoir at large $k_y$. This may provide an efficient relaxation channel for the excitations to reach the bottom of the polariton dispersion (in $k_y$, which is the top of the dispersion in $k_x$, instead), ultimately allowing for BEC formation, as  shown in the following.}

\begin{figure}
    \centering
    \includegraphics[width=\textwidth]{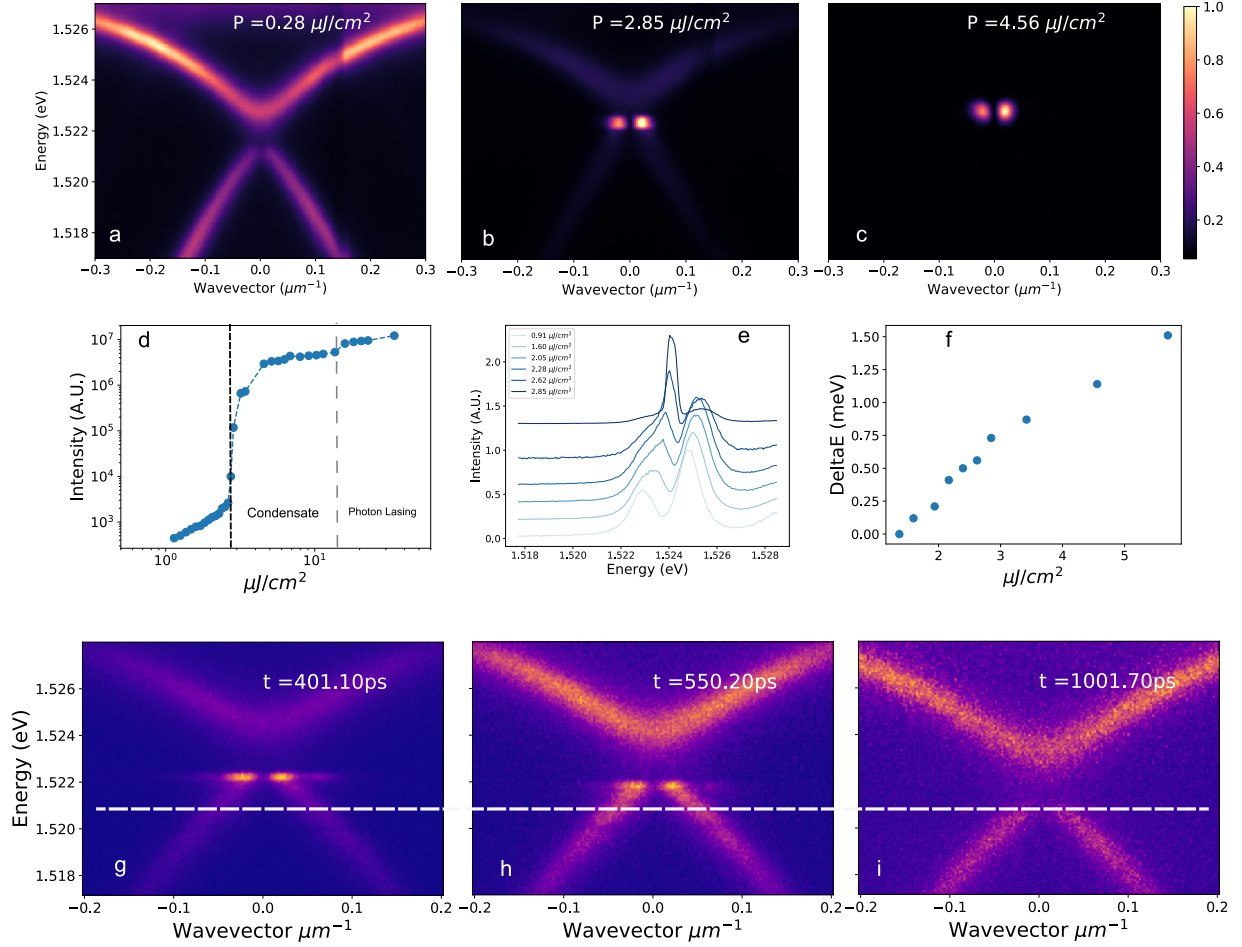}
    \caption{
    \textbf{Polariton BIC condensation.} (\textbf{a-c}) Angular resolved PL emission under  non-resonant pulsed (ps) excitation; the excitation power is increased from a to c. In panel b a double peaked emission around k$\sim 0$ emerges; when the pumping power is increased this emission becomes dominant as shown in (c). (\textbf{d}) Output intensity of the double peak emission as a function of exciting energy density, showing a clear threshold behaviour and demonstrating phase transition to a condensed state in the BIC mdoe. (\textbf{e}) Characteristic coherence buildup at threshold, resulting in the narrowing of the condensate linewidth. (\textbf{f}) Blueshift of the polariton emission in the power range around condensation threshold. (\textbf{g-i}) Temporal and angular resolved PL emission after the non-resonant pulsed excitation for a pumping power close to the condensation threshold. The snapshots, taken after $401$, $550$ and $1000$ ps, respectively, show that the condensate emission always comes from the BIC on the top of the lower branch despite the time-dependent blueshift. The full movie showing the complete dynamical evolution of the PL can be accessed as a supplementary video file.}
    \label{Condensate}
\end{figure}

Next, we address the nonlinear properties of the polariton BIC occurring at the dispersion saddle point, showing clear evidence of polariton condensation. Experimental results obtained by pumping with a non-resonant pulsed laser, focused on one of our etched gratings, 
are shown in Fig.~\ref{Condensate}. The low power dispersion in Fig.~\ref{Condensate}a is similar to the one in Fig.~\ref{sketch}d. 
On increasing the pumping power (from panel a to c) the measured polariton dispersion shows the appearance of a two-lobe emission, becoming dominant at higher powers. This behaviour is characterised by a nonlinear increase of the two lobes intensity,  as shown in Fig.~\ref{Condensate}d, a reduction of the linewidth (Fig.~\ref{Condensate}e), and a blue shift (Fig.~\ref{Condensate}f), typical of a phase transition to a polariton BEC. The strong blue-shift as a function of pump power, which is due to polariton-polariton repulsive interaction, confirms that such coherent state emission is not given by conventional photon lasing in the weak coupling regime.
Although a net transition is hard to identify under pulsed excitation, at extremely high powers a second threshold can be noticed (Fig.~\ref{Condensate}d), which eventually identifies the transition to the weak coupling regime. We further notice that the threshold of polariton BEC is observed at a pump 
corresponding to an energy density per pulse of the order of 3 $\mu$J/cm$^{2}$. These experiments demonstrate that the double-peaked emission comes from a polariton condensate,\cite{Kasprzak2006} leaking out (acquiring finite momentum) from the polariton BIC (which is dark at exact normal incidence). 
In Fig.~\ref{Condensate}(g-i) we report a few typical snapshots obtained from time-resolved streak camera measurements at a pumping power close to the condensation threshold. These panels clearly show how the condensate emission always comes from the polariton BIC mode, even if a strongly time-dependent shift of the emission is observed. Moreover, the condensate linewidth in the time-integrated images appears slightly blurred due to this dynamical change of energy, which is evidenced in the time resolved panels. Finally we would like to notice that 
no mixture between the BIC  and the bright mode at the minimum of the upper branch is observed.    

\begin{figure}[!ht]
    \centering
    \includegraphics[width=\columnwidth]{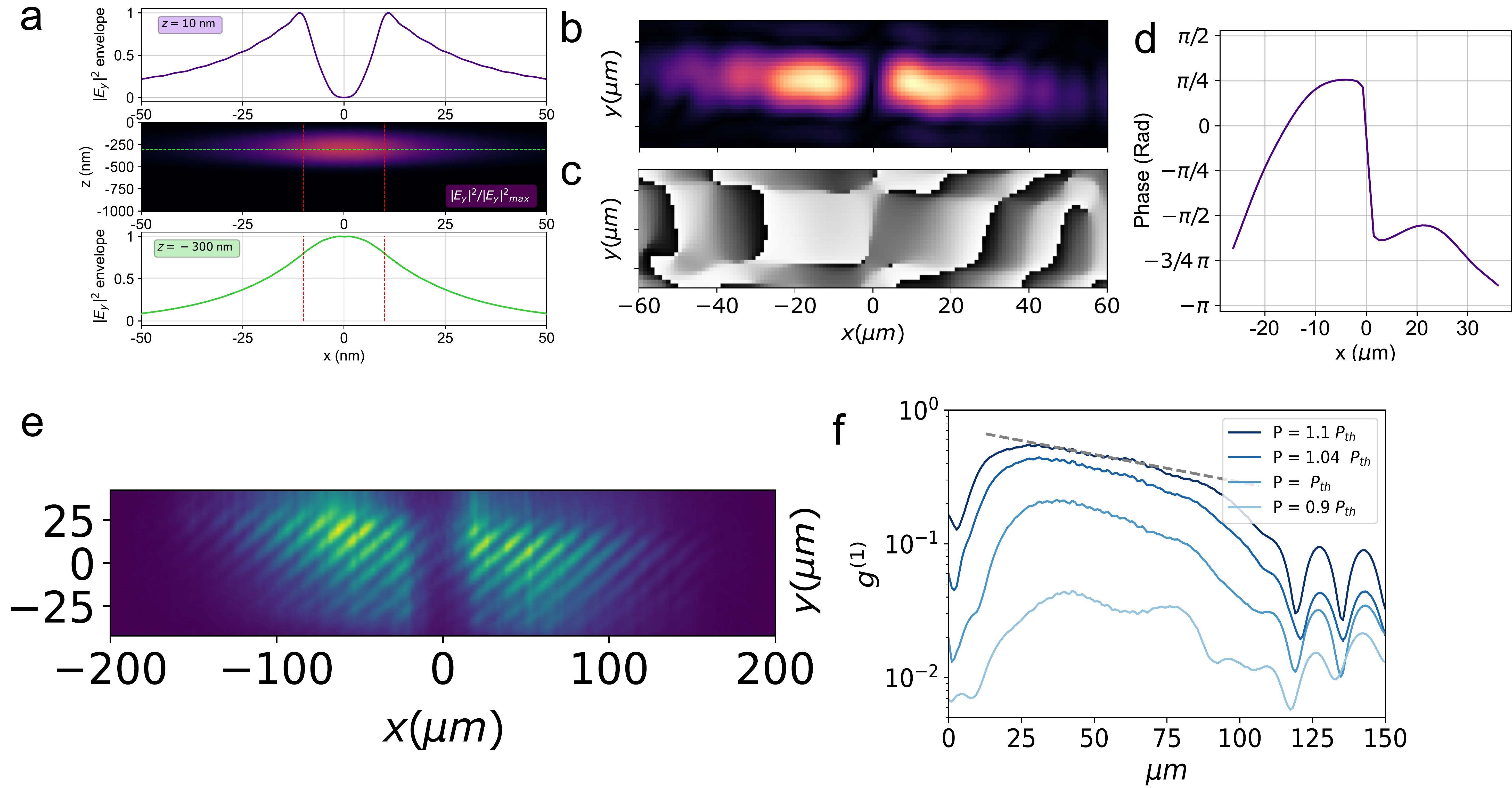}
    \caption{\textbf{Coherence and phase singularity of polariton BIC condensate.} (\textbf{a}) Numerical simulation of photonic BIC mode shallow confinement along the grating periodicity direction: the laser spot produces a local enhancement of exciton population, which can be interpreted as a lowering of the effective index, hence a local blue shift of the BIC mode, similar to the heterostructure confinement in a QW; in the middle the electric field intensity of the BIC mode is plotted as a function of $x$ and $z$, in the bottom panel the envelope of this solution is reported, corresponding to the position indicated by the dashed green line in the middle, i.e. within the waveguide core, and finally the top panel shows the spatial emission of the polariton BIC from the top of the structure (i.e., immediately above the patterned grating). 
    (\textbf{b}) Experimentally measured spatial emission of the condensate from the polariton BIC, and (\textbf{c}) phase pattern corresponding to (b); (\textbf{d}) Measured phase profile showing the $\pi$-shift of the phase between the two spatial lobes of the emission profile in panel (b), resulting in the destructive interference around $x\sim 0$.
    (\textbf{e}) Interferogram of the polariton condensate from the polariton BIC, and (\textbf{(f)}) first-order correlation function for different pumping powers across the condensation threshold; by fitting the exponentially decaying portion of the correlation function, we extract a coherence length as high as 200 $\mu$m at $P \sim P_{th}$.}
    \label{Phase}
\end{figure}

The BEC distribution in space is strongly affected by the anti-symmetric modal properties inherited from the purely photonic BIC on the grating  periodicity scale, while it acquires a slowly varying envelope due to the pump spot, as simulated by FDTD (neglecting the excitonic resonance) and shown in Fig.~\ref{Phase}a. This is evidenced by filtering the high-frequency oscillating components from the whole calculated mode. The envelope of the squared modulus of the relevant electric field component is plotted within the core region of the effective waveguide, i.e. below the patterned region. However, when plotting the envelope a few nm above the grating surface, this shows a peculiar anti-symmetric profile (top panel). Remarkably, this compares very well with the near-field image of the measured intensity emitted from the polariton condensate in Fig.~\ref{Phase}b.
We also experimentally extract the phase map of the polariton BIC emission. This is done by taking a small portion of the emission as a phase reference, and then letting this reference interfere with the whole spatial emission. The resulting interference pattern is shown in Fig.~\ref{Phase}c. A $\pi$ phase jump is visible across the dark state, as it is also highlighted by the horizontal cross-section shown in Fig.~\ref{Phase}d. This is consistent with the calculated far-field phase profile for the purely photonic BIC mode (see SM).\\
The spatial coherence of the polariton condensate can be obtained from the interferogram shown in Fig.~\ref{Phase}e. The interferogram is obtained by splitting the image of the emission in a Michelson interferometer and spatially flipping the image of one of the two arms (see SM for details). 
The first-order correlation function is obtained from the fringe visibility and shown in Fig.~\ref{Phase}f, demonstrating that the condensate is coherent almost over the entire grating length. 

\begin{figure}
    \centering
    \includegraphics[width=\columnwidth]{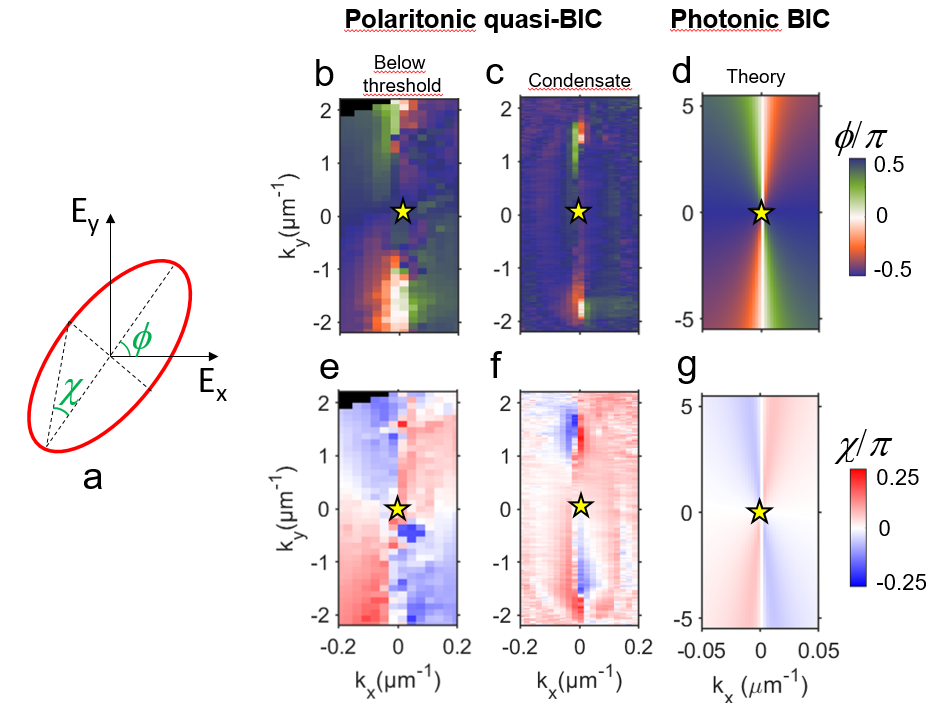}
    \caption{\textbf{Polarization vortex of polariton BEC in a BIC.} 
    (\textbf{a}) Sketch of the polarization ellipse parameters: $\phi$ represents the polarization vector direction, while $\chi$ represents the ellipticity or degree of circular polarization. (\textbf{b}) $\phi$ and (\textbf{e}) $\chi$ parameters measured for the polariton BIC branch below condensation threshold; (\textbf{c}) $\phi$ and (\textbf{f}) $\chi$ parameters measured for the condensate from the polariton BIC above threshold; (\textbf{d}) $\phi$ and (\textbf{g}) $\chi$ parameters calculated for the purely photonic BIC branch; The agreement between the data and the simulation shows that the polariton condensate from the polariton BIC inherits the nontrivial topological properties of the photonic BIC.}
    \label{Polarization}
\end{figure}

Finally, in Fig.~\ref{Polarization} we address the topological properties acquired by the polariton BIC condensate, by directly measuring the polarization vortex in far-field. As it has already been discussed, photonic BICs possess topological charges that are directly related to the polarization pattern in the momentum space of the band hosting the BIC mode \cite{Zhen2014, Doeleman2018, Zhang2018}. We have characterized the polarization pattern emission in the far-field both below and above the condensation threshold by measuring the parameters of the polarization ellipse, i.e., the polarization direction $\phi$ and the ellipticity $\chi$, as defined in Fig.~\ref{Polarization}a. Figures \ref{Polarization}b,e  show the polarization parameters measured for the polariton BIC branch below condensation threshold. A polarization vortex is clearly visible, confirming that the hybrid light-matter polariton BIC studied in this work inherits the topological nature of the {purely photonic} BIC whose polarization properties are computed in Fig.~\ref{Polarization}d,g. 
Most important, the same polarization pattern is measured also from the BEC emission (panels c and f), even if the latter is concentrated in a much smaller region of the reciprocal space with respect to the whole polariton BIC branch. These results show that the condensation from a polariton BIC can be used as an alternative route to imprint a topological charge to a BEC.

In summary, we have exploited bound states in the continuum to enhance the lifetime of radiative excitations in a polariton waveguide, which allowed to achieve polariton Bose-Einstein condensation at low power density threshold. Thanks to the high quality factor of the polariton BIC, condensation takes place even without an absolute energy minimum in the dispersion: by directly accessing the system dispersion through angle and energy resolved PL measurements, we have observed that the BIC in a simple waveguide-grating system is part of a more complex, saddle-like dispersion.
Our work also opens a promising route for controlling the polariton condensate properties in a new way, i.e., by transferring topological properties from a photonic structure\cite{Zhen2014}  to a macroscopic quantum fluid of light, with potential applications to metasurface exciton-polaritons in alternative material platforms \cite{Deng2018_ncomm,HaMy2020,Majumdar2020} using simple shallow patterning of planar waveguides.    

\vspace{1cm}
\noindent \textbf{Methods}\\

\noindent \textit{Sample design and fabrication.} 
{The semiconductor heterostructure considered in this work is an effective planar waveguide, whose core is made of a periodic repetition of 12 (20 nm thick) QW and (20 nm thick) barrier layers, as schematically represented in Fig.~\ref{sketch}a. The first 2 QWs are partially etched to create a 1D grating structure. The total thickness of the waveguide core is about 500 nm (plus the 10 nm-thick capping GaAs layer), the etched grooves are 90 nm deep, and the lattice constant is taken as $a=243$ nm, which allows to bring the relevant photonic mode in resonance with the QW excitonic transitions, when folded around $k_x=0$ in reciprocal space. The grating is extended for up to 300 $\mu$m along the direction normal to the grooves and  50 $\mu$m along the direction parallel to the grooves. Photonic mode dispersions (see Fig.~\ref{sketch}) are calculated 
with the finite difference time domain method (FDTD, see details in the SM file). 
In the realized sample, the two photonic modes of Fig. \ref{sketch}b strongly couple to the excitonic transition (centered at $E_{X}\sim 1.527 eV$), with a Rabi splitting of about 13 meV. Due to symmetry considerations the coupling and the resulting eigenmodes can be described by a model of 4 coupled oscillators (see SM for details). The resulting polariton dispersions are shown in Fig. \ref{sketch}c. \\
The fabrication of the actual grating structures was performed by spin coating a positive electron beam sensitive resist (ZEP-520a) on the sample. The resist was exposed using Vistec VB300 Electron Beam Lithography System, and then developed in amyl-acetate. The etching was performed using an Oxford PlasmaLab 150 Inductively Coupled Etcher, tuning the recipe to induce low damage in the sample. Finally, a post-processing of hydrochloric acid bath and subsequent 8nm-thick conformal Al$_{2}$O$_{3}$ layer were used to passivate the surface defects induced during the etching.
}

\noindent \textit{Measurements.} All the optical measurements are realized at cryogenic temperatutre (4K). The energy-momentum polariton dispersion is accessed by angular and energy resolved photoluminescence (PL) emission. A typical example is shown in Fig.~\ref{sketch}d for a measured sample with a grating having a pitch $a \sim 240$ nm and a $FF\sim 0.7$. We use a streak camera to reconstruct the temporal dynamics of the dispersion renormalization and polariton condensation as it is shown in Fig.~\ref{Condensate}. In particular, these panels correspond to the highest observed blueshift at a given pumping powers. See also SM and the online videos for further details. The pumping spot is circular or elliptical with a FWHM of the ellipse major axis $S$ ranging from almost 200 $\mu m$ down to 30 $\mu m$ (see also measurements reproduced in the SM). Finally, the coherence properties are extracted by using a Michelson interferometer to retrieve both the phase map and the interferograms. See also SM for further details.

\noindent \textbf{Acknowledgments}\\
We thank Paolo Cazzato for technical support.\\
We are grateful to Lucio C. Andreani for useful discussions and support, and Ronen Rapaport for inspiring discussions and for sharing information about the sample design.\\
The authors acknowledge the Italian Ministry of University (MUR) for funding through the PRIN project ``Interacting Photons in Polariton Circuits'' – INPhoPOL (grant 2017P9FJBS). \\
Work at the Molecular Foundry was supported by the Office of Science, Office of Basic Energy Sciences, of the U.S. Department of Energy under Contract No. DE-AC02-05CH11231.\\
We thank Scott Dhuey at the Molecular Foundry for assistance with the electron beam lithography.\\
We acknowledge the project FISR - C.N.R. “Tecnopolo di nanotecnologia e fotonica per la medicina di precisione” - CUP B83B17000010001 and "Progetto Tecnopolo per la Medicina di precisione, Deliberazione della Giunta Regionale n. 2117 del 21/11/2018.\\
This research is partly funded by the Gordon and Betty Moore Foundation’s EPiQS Initiative, Grant GBMF9615 to L. N. Pfeiffer,  and by the National Science Foundation MRSEC grant DMR 1420541.\\
H.S.N is partly supported by the French National Research Agency (ANR) under the project POPEYE (ANR-17-CE24-0020) and the IDEXLYON from Université de Lyon, Scientific Breakthrough project TORE within the Programme Investissements d’Avenir (ANR-19-IDEX-0005).\\ 
\noindent \textbf{Competing Interest}\\
The authors declare no competing interest.\\
\noindent \textbf{Author Contribution}\\
V.A., D.S. and D.G. initiated the research project, V.A. and F.R. designed the grating structures with inputs from H.S.N; F.R. processed the sample, growth was performed by K.B and L.P.; V.A, F.R., A.G., D.G.S.F., M.E.T., H.S.N and D.B. realized the experiments and carried out the data analysis; H.S.N, S.Z. and D.G. developed the theoretical background and performed the FDTD simulations. V.A, H.S.N, D.G., and D.S. drafted the manuscript, and all the authors were involved in the discussion of results and the final manuscript editing.

\bibliographystyle{unsrt}
\bibliography{biblio.bib}

\newpage

\section{Supplementary Material file to: \\ Polariton Bose-Einstein condensate from a Bound State in the Continuum}

\subsection{Finite Difference Time Domain simulations of the passive heterostructure}

Finite difference time domain (FDTD) simulations have been employed to calculate both the photonic modes and their dispersion, as well as far field profiles. 
This method provides the electrodynamics solution of Maxwell equations for arbitrary complex systems, by discretizing and solving the finite difference formulation of the coupled differential equations \cite{Taflove2005}. Throughout the work, we have used a commercial software implementation of the FDTD algorithm from Lumerical-Ansys (https://www.lumerical.com/ansys/).
We notice that only the purely photonic heterostructure is considered here, i.e., without including any excitonic resonance in the dielectric response but assuming a dispersionless refractive index for either the quantum well or the barrier layers. This allows to quantitatively derive the BIC-like properties of the grating modes, to be then coupled to the excitonic degrees of freedom in a coupled oscillators model.

\begin{figure}
    \centering
    \includegraphics[width=0.98\columnwidth]{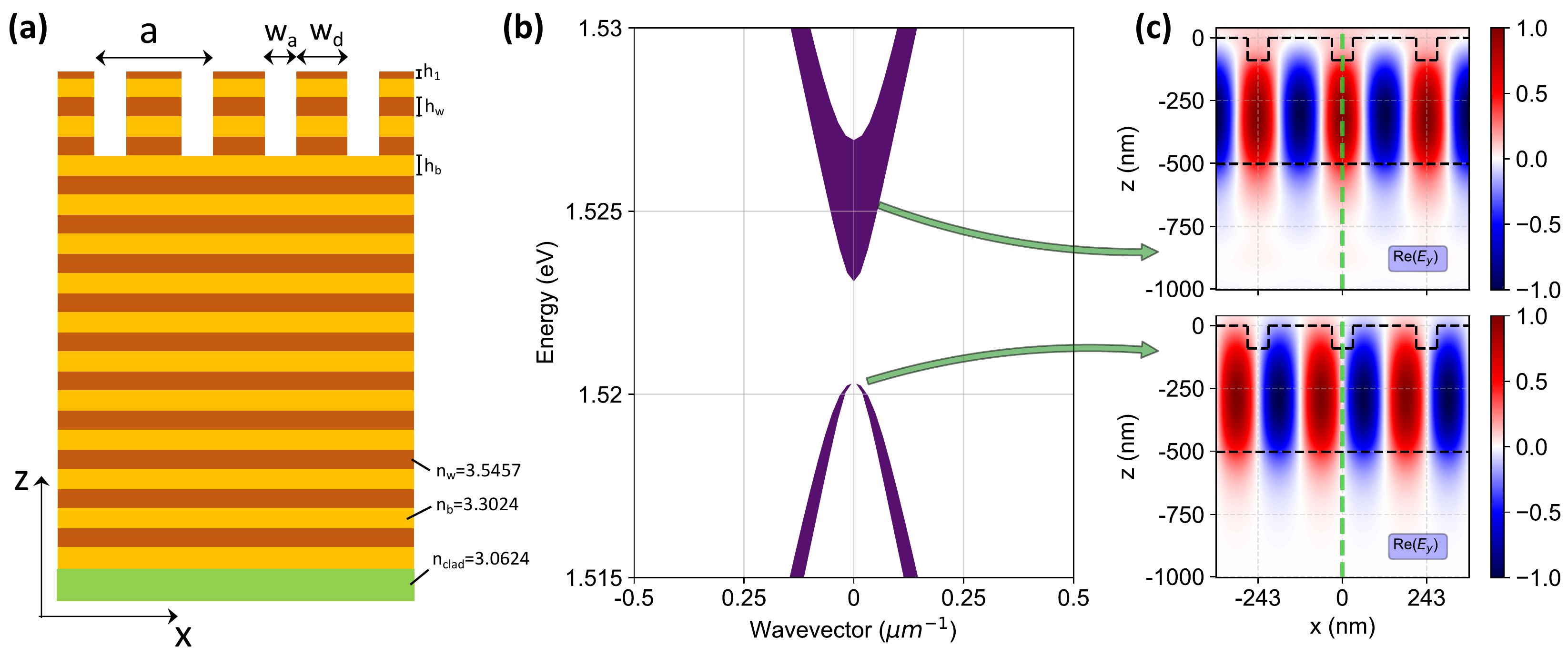}
    \caption{(a) Sketch of the simulated heterostructures, corresponding to the fabricated devices, and definition of geometric parameters. The lower cladding is assumed to be a 500 nm-thick layer sitting on a GaAs substrate, which is simulated as a semi-infinite layer by imposing PML boundary conditions. (b) Simulated mode dispersion of the grating modes around normal incidence ($k=0$), assuming structure parameters as follows: $a=243$ nm, $w_a=60.75$ nm (i.e., $w_a/a=0.25$), cap layer $h_1=10$ nm, barrier layer $h_b=20$ nm, QW layer $h_w=20$ nm, etch depth 90 nm. The line thickness is coded to represent the k-dependent Q-factor of each mode (i.e., proportional to $1/Q$), thus highlighting the dark mode at $k=0$. (c) Electric field profiles for the TE component, as numerically obtained for either the symmetric (i.e., bright) mode or the antisymmetric (i.e., dark, or BIC) mode, as it is clearly evident by placing the center of the grating unit cell in $x=0$. The simulated profile of the grating structure is superimposed with a dashed line. }
    \label{FDTD_sims}
\end{figure}

\textit{Grating simulations}.\\
The photonic mode dispersion of complex dielectric heterostructures with periodic repetition of an elementary cell along one or more spatial directions can be obtained by exciting the system with a collection of broad-band, point-like electric dipole sources that are randomly placed in the whole elementary cell. This allows to excite all the possible resonant modes in the structure, and proper phases are given to each dipole in order to satisfy the Bloch boundary conditions for given in-plane wave vectors; thus, each wave vector in the first Brillouin zone of the corresponding photonic lattice requires a separate FDTD simulation, and by running over all the k-vectors, the whole band structure is reconstructed (see, e.g., the grating bands reported in Fig. 1 of the main text). Notice that the perfectly matched layer (PML) boundary conditions allow to simulate an infinitely extended dielectric medium (e.g., in the upper and lower claddings), and the resonant modes of the structure are calculated as the ones that retain most of the electromagnetic energy density within the simulation region. Each mode falling within the light cone of the vertical heterostructure is intrinsically lossy in the claddings, which is the source of finite linewidths (even in the absence of further losses introduced, e.g., by exciton absorption). This `imaginary part' of photonic modes is directly obtained in FDTD by fitting the exponential decay of the fields associated with each specific resonant mode excited in the simulation region. This exponential decay in time is associated to the mode Q-factor, from which its intrinsic loss rate can straightforwardly be defined as $\gamma=\omega/Q$, where $\omega$ represents the center mode frequency. A Q-factor larger than $10^6 - 10^7$ is typically associated to a BIC, for these structures. The linewidth of the photonic modes dispersion plotted, e.g., in Fig. 1 of the main text is proportional to $\gamma$. Specifically, the line thickness in Fig. 1 is coded to be inversely proportional to $Q_k$, where $Q_k$ is the mode Q-factor at each wave vector in the first Brillouin zone defined by the grating periodicity (in particular, we used $15/Q_k$ for the lower branch, and $10/Q_k$ for the upper branch). 

\begin{figure}[t]
    \centering
    \includegraphics[width=0.92\columnwidth]{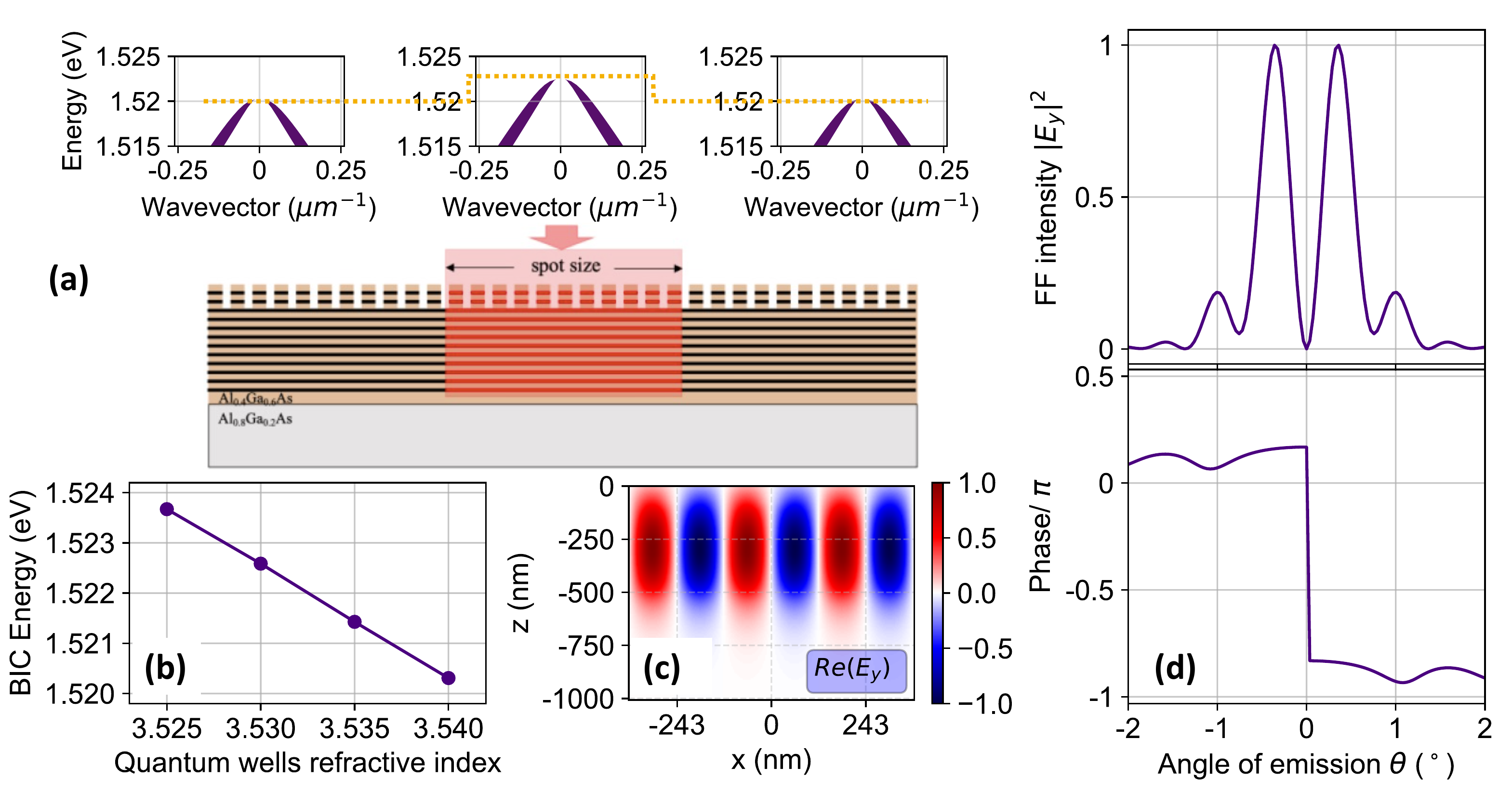}
    \caption{(a) Schematic representation of the confined quasi-BIC mode of the grating: the laser spot produces a local enhancement of exciton population, which can be interpreted as a lowering of the effective index, hence a local blue shift of the polariton BIC mode; similarly to the heterostructure confinement in a QW, this creates an effective confining potential for the polariton BIC. (b) Dependence of the BIC energy in the grating mode at $k=0$ calculated by FDTD and plotted as a function of the GaAs refractive index in the 12 QW layers of the heterostructure. (c) Real part of the transverse electric field component plotted over 3 lattice cells of the structural grating around its center, simulated for the whole 150 $\mu$m-wide grating structure in which the central 20 $\mu$m-wide region is assumed with a lower refractive index within the 12 QW layers. (d) Far field projection of the surface near-field simulated in panel (b), plotted as a modulus squared of the transverse field component (upper panel); the corresponding phase is plotted in the lower panel. }
    \label{Spotsize_th}
\end{figure}

It is worth pointing out that FDTD is computationally costly. However, we have opted for this approach since it virtually provides a numerically exact solution of the electromagnetic problem in such as complex dielectric heterostructure, i.e., with sufficiently accurate spatial discretization the mode dispersion, losses, and electromagnetic field distributions should converge to the exact solutions. For the particular case of the electromagnetic simulations reported in this work, we have chosen to simulate dispersionless dielectric media, identified with refractive indices corresponding to the background values of the given compound. In particular, here we  assume $n_w\sim 3.54$ for the GaAs layers, $n_b\sim 3.3$ for the Al$_0.4$Ga$0.6$As, and $n_{clad}\sim 3.06$ for Al$0.8$Ga$0.2$As cladding, respectively. The relevant structure parameters are defined in the sketch reported in Fig.~\ref{FDTD_sims}(a), according to the notation reported in the main text. The specific values employed in the simulations reported here are given in the caption. The overall waveguide core thickness is about 500 nm. In terms of real space meshing, the photonic bands of the purely periodic grating have been simulated with $\Delta z=2.5$ nm and $\Delta x=1.5$ nm as the mesh steps along the growth and periodicity directions, respectively, previously tested for convergence. The pulse length of the emitting dipole sources has been set to $t_{\text{pulse}}\approx 10$ fs, which implies the broadband emission spectrum necessary for the band structure calculations. In order to avoid spurious resonances, the decaying fields are recorded starting from $t_i=100$ fs up to $t_f=8000$ fs. Results are reported in Fig.~\ref{FDTD_sims}(b), where the same modes dispersion around normal incidence as the one in the inset of Fig. 1 in the main text is here reproduced for completeness. In the panels of Fig.~\ref{FDTD_sims}(c) we also plot the steady state electric field profiles calculated for the bright (upper panel) and the dark (lower panel) $k=0$ modes, respectively. While the former is clearly symmetric with respect to the center of the unit cell in $x=0$, the latter is antisymmetric, which confirms the BIC nature of the mode. These field profiles are calculated by using much longer excitation pulses, in order to spectrally isolate the required resonances and extracting the fields more precisely. \\

\textit{Laser spot as an effective confining potential}.\\
An enhancement of the exciton population causes a local lowering of the QWs refractive index. As a consequence, a local blueshift of the whole photonic band structure is expected. In fact, the non-resonant pulsed laser employed in the experiments reported in this work creates a large free carriers density, which produces a local blueshift caused by the interaction of polaritons and the reservoir (i.e., a refractive index decrease within the multi-QW heterostructure). This can be schematically represented as in Fig.~\ref{Spotsize_th}(a), with a local blue-shift of the band structure folded by the grating periodicity around normal incidence, and in particular the polariton BIC mode at $k_x=0$.
The BIC energy for the grating structure at $k=0$ is plotted in Fig.~\ref{Spotsize_th}(b), as a function of the refractive index in the GaAs layers.
Hence, for the laser spot-induced confining potential, we have simulated the effect of the local nonlinear shift under the exciting beam by modifying the GaAs refractive index to $n=3.53$ within a region of lateral width $w_c=20\, \mu$m. This is enough to produce a blue-shift of the BIC mode in the order of 2 meV, consistent with the experimental evidence.
The non-resonant pumping then creates an effective 1D well for a negative mass particle, which produces a localized quasi-BIC mode along the periodicity direction of the grating, as simulated by FDTD. This is similar to the heterostructure confinement of positively charged holes in semiconductor quantum wells, where the negative mass electronic band extreme (top of the valence band for electrons) is spatially dependent and it allows to confine electron vacancies (i.e., holes) within the layer with highest energy band extreme. Here, the confined excitations can be interpreted as negative mass polaritons. \\
In the FDTD simulations, we have directly simulated a $150\,\mu$m wide sample. In this case, the mesh steps along the two directions are $\Delta z=5$ nm and $\Delta x=6$ nm, respectively. In order to selectively excite only the fundamental cavity mode, we employed only one point-like source located in an antinode of the electric field. Moreover, to help the selective excitation of the confined BIC mode, a narrow-band dipole source is used, for which we set the pulse length to $t_{\text{pulse}}=900$ fs, with the emission peak centered at $\lambda = 816$ nm. The field profiles have been recorded with specific monitors tuned at $\lambda = 816$ nm starting from $t_i=10000$ fs. Since the field profiles are recorded along the whole $150\,\mu$m wide structure, and the rapid oscillations of the electric fields occur on the lattice constant ($a=0.243$ $\mu$m) scale, it is useful to retrieve only the slowly varying envelopes of the fields. These envelopes can be obtained by filtering all the high-frequency components in their spatial Fourier spectrum. The low frequency components are the ones corresponding to the slowly varying envelopes that are plotted in Figure 3 of the main text. To confirm that the fundamental mode of the negative mass trap potential is actually a quasi-BIC, we plot in Fig.~\ref{Spotsize_th}(c) the corresponding electric field profile in steady state, on the spatial scale corresponding to the lattice periodicity of the structural grating. Shortly, this is a close up of the full intensity profile reported in Fig. 3 of the main text. Here, by explicitly plotting the real part of the field, its antisymmetric nature is clearly evident (as it can be directly inspected from a comparison with the lower panel of Fig. ~\ref{FDTD_sims}(c)). Finally, in Fig.~\ref{Spotsize_th}(d) we can appreciate the far field properties of the calculated near field shown in ~\ref{Spotsize_th}(c). In particular, the surface electric field taken a few nm above the grating surface is near-to-far field projected, resulting in the top panel with an intensity node at normal incidence. Correspondingly, this far field intensity profile is associated with the phase discontinuity reported in the lower panel of the Figure. It is worth noticing the close similarity of the latter plot with the experimentally measured phase discontinuity reported in Fig.~3 of the main text for the far field emission of the polariton BIC.\\
As a final comment, we notice that the actual presence of this shallow 1D confining potential is confirmed from the experimental data reported in the following on varying spot sizes, showing the different orders of the longitudinally confined modes (see, in particular, Fig.~\ref{Spotsize} and the related paragraph).

\section{Analytic theory of quasi-BIC polaritons in waveguide gratings}

\subsection{Two-step perturbation method}

In reference \cite{Lu2020}, Lu \textit{et al.} have developed an analytic model for the strong coupling regime between excitonic resonances and photonic BIC in 1D lattice. Here, we will expand the theory of Lu \textit{et al.} to describe the polaritonic dispersion along both $k_x$ and $k_y$ directions, as well as the polarization texture of polaritonic bands, which is relevant to the understanding of data presented in the main text.\\

\begin{figure}
	\begin{center}
	\includegraphics[width=1 \textwidth]{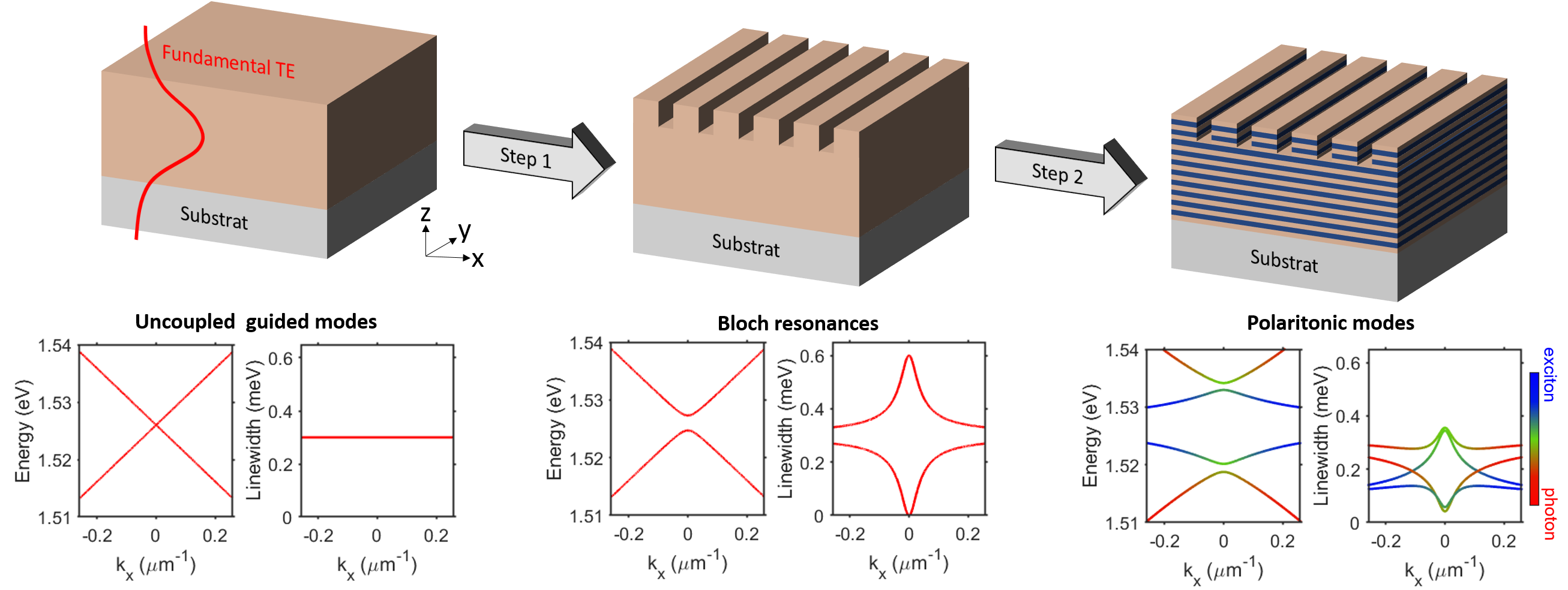}	
\caption{Sketch of the two-step perturbation method.}
	\label{fig:2step}
\end{center}
\end{figure}

We focus on the fundamental TE (Transverse Electric) guided mode and its modification due to the periodic corrugation, as well as the strong coupling effect with an excitonic resonance, in a two-step perturbation approach (see Fig.\ref{fig:2step}):
\begin{itemize}
    \item \textbf{\textit{Step 1}}: The interaction between guided modes thanks to Bragg scattering coupling and the radiative continuum, resulting in Bloch resonances exhibiting a gap opening and the formation of symmetry-protected Bound state In the Continuum (BIC) at the $\Gamma$ point of the momentum space.
    \item \textbf{\textit{Step 2}}: The strong coupling regime between Bloch resonances and excitonic resonances to make Bloch exciton-polariton modes, in particular the formation of quasi-BIC when hybridizing photonic BIC and excitons. 
\end{itemize}

\subsection{Band folding for uncoupled guided modes}
We first discuss the band folding effect of the 1D photonic grating of period $a$, corrugated along $x$-axis. A given $(k_x,k_y)$ point in the vicinity of the first $\Gamma$ point can be attributed to two  propagation vectors, given by: 
 \begin{equation}\label{eq:beta_1}
 \vec{\beta_{\pm1}}=\left(k_x \pm\frac{2\pi }{a}\right)\vec{u_x} + k_y\vec{u_y}.
 \end{equation}
 with $\vec{u_x}$ and $\vec{u_y}$ the unit vectors along $\hat{x}$ and $\hat{y}$ directions.\\
 
Concentrating on the first $\Gamma$ point, \ul{\textit{the basis for the coupling mechanisms in Step 1 of our perturbation theory is constituted of two corresponding guided modes: $\ket{\vec{\beta_{-1}}}$ and $\ket{\vec{\beta_{+1}}}$}}. The dispersion of these states can be approximated as a linear dispersion having group velocity $v_g=c/n_g$ with $n_g$ being the group index and $c$ being the speed of light. Thus the dispersion characteristic $\omega_{\pm1}(k_x,k_y)$ corresponding to $ \vec{\beta_{\pm1}}$ is given by:
 \begin{align}
 \begin{split}
     \omega_{\pm1}(k_x,k_y)&=\omega_0 + \frac{c}{n_g}\left(\lvert \vec{\beta_{\pm1}} \rvert-\frac{2\pi}{a}\right)\\
     &\approx \omega_0 \pm \frac{c}{n_g}k_x + \frac{ca}{4\pi n_g}k_y^2\label{eq:dispersion_origin}
     \end{split}
 \end{align}
 where $\omega_0$ is the pulsation of the guided mode at $\Gamma$ point. 
 To work with dimensionless quantities, we redefine the wavevectors, as well as the pulsations as:
\begin{align}
 \begin{split}
 q_x&=k_xa/2\pi, \\
 q_y&=k_ya/2\pi, \\
 \hat{\omega}_{\pm1} &= \left(\omega_{\pm1}-\omega_0\right)a/2\pi c.
 \end{split}
\end{align}
the dispersion characteristic \eqref{eq:dispersion_origin} is then rewritten as:
\begin{equation}
 \hat{\omega}_{\pm1}(q_x,q_y)\approx \pm \frac{q_x}{n_g} + \frac{q_y^2}{2n_g}.
 \end{equation}
 The electric field $\vec{E_{\pm1}}$ of $\ket{\vec{\beta_{\pm1}}}$ is in-plan and perpendicular to the corresponding propagation vector $\beta_{\pm1}$ . Thus its polarization vector $\vec{u_{\pm1}}$  satisfies $\vec{u}_{\pm 1} \cdot \vec{\beta}_{\pm 1}=0$ is given by:
 \begin{equation}
    \vec{u_{\pm1}}(q_x,q_y) = \cos\theta_{\pm1}\vec{u_x} + \sin\theta_{\pm1}\vec{u_y},\label{eq:u_pm1}
\end{equation}
    with     
    
    \begin{align}
    \begin{split}
    \cos\theta_{\pm1}&=-\frac{q_y}{\sqrt{(q_x\pm 1)^2 + q_y^2}}\\ \sin\theta_{\pm1}&=\frac{q_x\pm 1}{\sqrt{(q_x\pm 1)^2+ q_y^2}}.\label{eq:theta}
    \end{split}
    \end{align}
Now the dispersion and polarization characteristics of our basis are well established. 

\subsection{Coupling between counter propagating guided modes and the formation of symmetry-protected BIC}
With only band folding effects, the dispersion of $\ket{\vec{\beta_{-1}}}$ and $\ket{\vec{\beta_{+1}}}$ crosses one to the other one. However, the introduction of an important periodic corrugation will induce a coupling, $U$, between $\ket{\vec{\beta_{-1}}}$ and $\ket{\vec{\beta_{+1}}}$. Moreover, being folded above the light line, each of them leak to the radiate continuum with the same coupling strength $\gamma_r$. In our basis $\Psi=\left( \ket{\vec{\beta_{+1}}},\ket{\vec{\beta_{-1}}}  \right)$, the photonic grating  can be described by the following effective Hamiltonian:
\begin{equation}\label{eq:H}
H(q_x,q_y)=\left(
\begin{matrix}
\hat{\omega}_{+1} & U\\
U & \hat{\omega}_{-1}
\end{matrix}
\right)
+ i\gamma_r \left(
\begin{matrix}
1 & \cos{\phi}\\
\cos{\phi} & 1
\end{matrix}
\right)
\end{equation}
The first term of $H$ describes the free oscillation of each mode and the coupling between them via diffractive coupling.  The second term of this Hamiltonian describes the radiative losses and the coupling between $\ket{\vec{\beta_{-1}}}$ and $\ket{\vec{\beta_{+1}}}$ via the radiative continuum. The coefficient $\cos{\phi}$ in the radiative coupling term is due to the fact that the loss exchange, i.e. farfield interference, is only for the same electric field component, thus given by $\vec{u_{-1}}.\vec{u_{+1}}=\cos{\phi}$, with:
\begin{equation}
 \phi(q_x,q_y)=\theta_{+1}-\theta_{-1}.   
\end{equation}
The eigenvalues $\Omega_\pm $ and corresponding eigenstates $\ket{\pm}$ of \eqref{eq:H} is given by:
\begin{equation}\label{eq:Omega}
    \Omega_\pm(q_x,q_y) = \frac{q_y^2}{2n_g} + i\gamma_r \pm \sqrt{\frac{q_x^2}{n_g^2} + \left(U-i\gamma_r\cos{\phi} \right)^2}.
\end{equation}
and
\begin{equation}
    \ket{\pm}=\ket{\vec{\beta_{+1}}} + C_\pm\ket{\vec{\beta_{-1}}}
\end{equation}
with
\begin{equation}\label{eq:C}
    C_\pm=-\frac{q_x}{n_g(U-i \gamma_r\cos{\phi})} \pm \sqrt{1+\left[\frac{q_x}{n_g(U-i\gamma_r\cos{\phi} )}\right]^2}.
\end{equation}
The real part $\hat{\omega}_\pm$ of $\Omega_\pm(q_x,q_y)$ corresponds to the normalized eigen-frequency. The imaginary counter part $\gamma_\pm$ corresponds to the radiative losses. For small $q_x, q_y\ll 1$, one may show that: 
\begin{subequations}
\begin{align}
\hat{\omega}_\pm (q_x,q_y)&\approx \frac{q_y^2}{2n_g} \pm U\left[1+\frac{q_x^2}{2n_g(U^2+\gamma_r^2)}\right],\label{eq:real}\\ 
\gamma_{\pm} (q_x,q_y)&\approx \gamma_r\left[1 \mp 1 \pm q_y^2  \pm \frac{q_x^2}{2n_g(U^2+\gamma_r^2)}\right].\label{eq:imag}
\end{align}
\end{subequations}

Interestingly, from \eqref{eq:imag}, we obtain directly \ul{\textit{$\gamma_{+} =0$ at $q_x=q_y=0$, and $\gamma_{+}$ increases as $q_x^2$ and $q_y^2$ when going out of the $\Gamma$ point. As consequence $\ket{+}$ is a BIC with an infinite quality factor at the $\Gamma$ point}}. On the other hand, the BIC nature of $\ket{+}$ at $q_x=q_y=0$ can also be explained from the symmetry of the eigenstate. Indeed, at $q_x=q_y=0$, Eq.~\eqref{eq:C} indicates that $C_\pm\rvert_{q_x=q_y=0}=\pm 1$, thus: 
\begin{equation}
    \ket{\pm_{q_x=q_y=0}}=\ket{\vec{\beta_{+1}}} \pm \ket{\vec{\beta_{-1}}}
\end{equation}
Moreover, the phase-shift $\pi$ between $\ket{\vec{\beta_{+1}}}_{q_x=q_y=0}$ and $\ket{\vec{\beta_{-1}}}_{q_x=q_y=0}$ suggests that $\ket{+_{q_x=q_y=0}}$ is an antisymmetric state while $\ket{-_{q_x=q_y=0}}$ is a symmetric one. \ul{\textit{Therefore, $\ket{+}$ cannot couple to the radiative continuum at $q_x=q_y=0$ due to symmetry mismatch and is a symmetry-protected BIC}}. This conclusion is in good agreement with the one deduced from the eigenvalues. 

\subsection{Farfield pattern, Polarization vortex and its associated Topological charge}
The farfield of the uncoupled guided modes $\ket{\vec{\beta_{\pm1}}}$ is  approximately a plane wave of wavevector $\vec{u_z}\omega_0/c$, with the farfield pattern $\vec{E_{\pm1}}(q_x,q_y)=E_0 \vec{u_{\pm1}}(q_x,q_y)$. Here $E_0$ is the field amplitude, and $\vec{u_{\pm1}}$ is given by \eqref{eq:u_pm1}. The hybridization of $\ket{\vec{\beta_{\pm1}}}$ into $\ket{\pm}$ corresponds to farfield patterns $\vec{E_{\pm}}$, with corresponding farfield patterns: 
\begin{equation}
    \vec{E_{\pm}}(q_x,q_y)=E_0\left(\vec{u_{+1}} + C_\pm\vec{u_{-1}}\right). \label{eq:farfield}
\end{equation}
Since $\vec{u_{\pm1}}(q_x=q_y=0)=\mp \vec{u_y}$ and $C_\pm(q_x=q_y=0)=\pm 1$, the farfield at $\Gamma$ point is simplified as:
\begin{align*}
\begin{cases}
    \vec{E_+}(q_x=q_y=0)&=0 \;\;\rightarrow \textit{ destructive interference}\\
    \vec{E_-}(q_x=q_y=0)&=-2E_0\vec{u_y}  \,\rightarrow\textit{ constructive interference}
    \end{cases}
\end{align*}
Therefore \ul{\textit{the farfield of $\ket{+}$ exhibits a singularity at $q_x=q_y=0$, as expected from its BIC nature.}}\\

\begin{figure}
	\begin{center}
	\includegraphics[width=0.5 \textwidth]{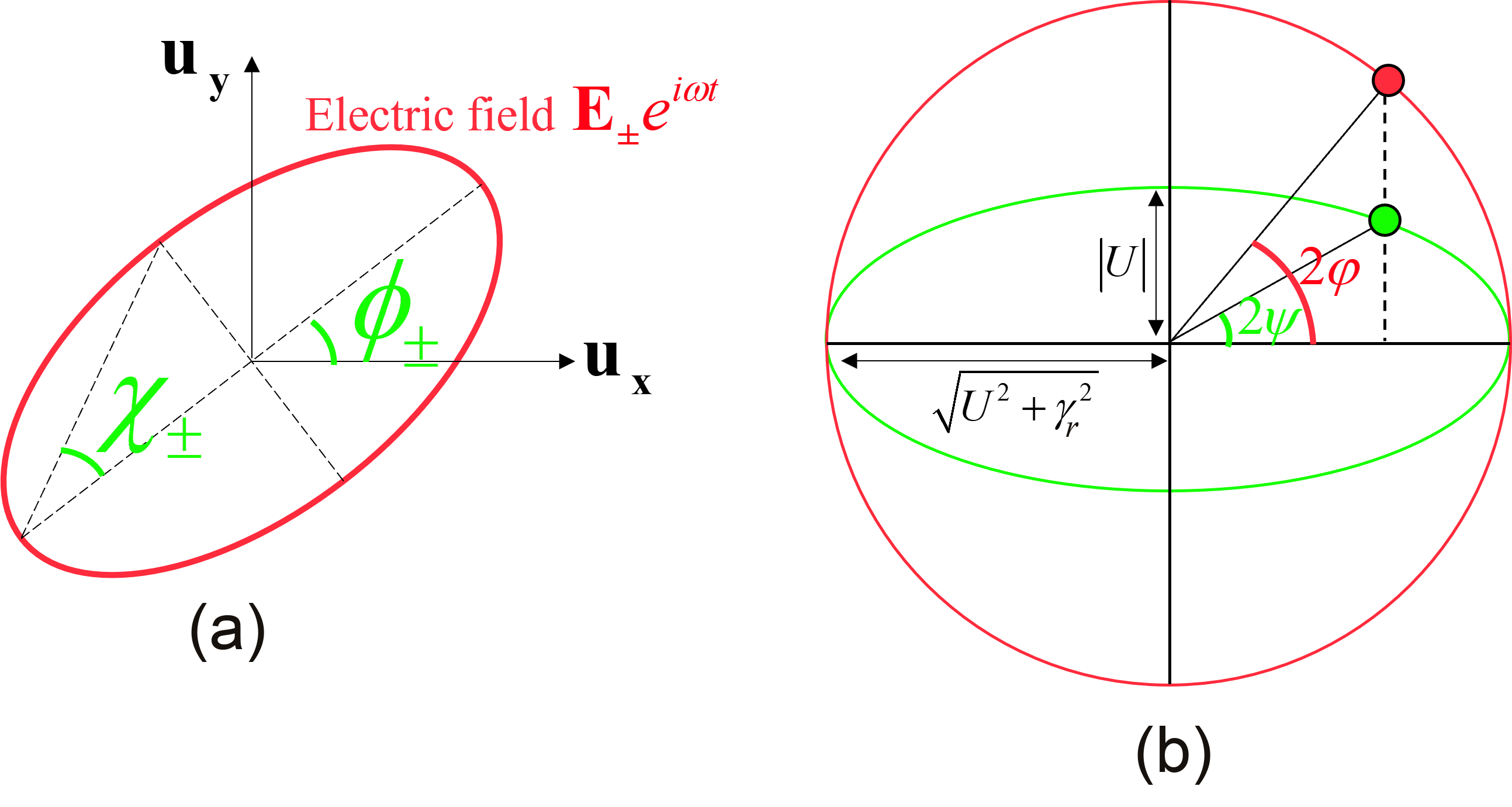}	
\caption{\textbf{(a):} Elliptic polarization of light with orientation angle $\phi$ and ellipticity $\chi$. \textbf{(b):} Graphical presentation of the 1:1 correspondence between $\varphi$ and $\psi$.}
	\label{fig:polarization_angle}
\end{center}
\end{figure}

\begin{figure}
	\begin{center}
	\includegraphics[width=0.8 \textwidth]{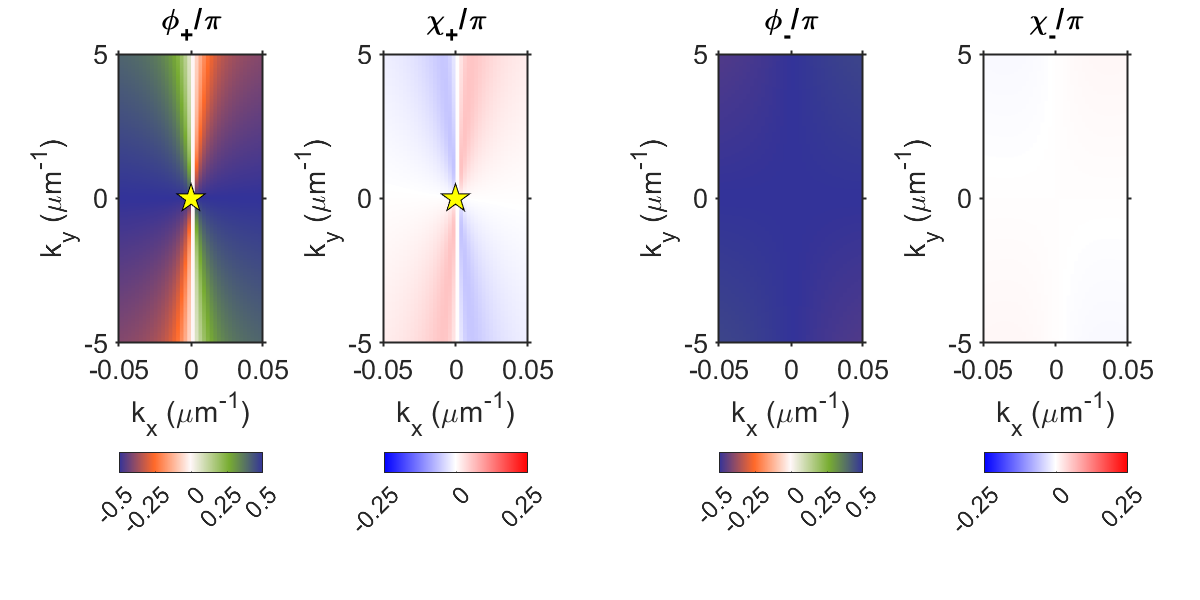}	
\caption{Mapping of the orientation angle and ellipticity of $\ket{+}$ (left), and $\ket{-}$ (right). Parameters: $a=243$ nm, $n_g=4$,  $\hbar \omega_0=1.526$ eV, $\gamma_r=10^{-4}$, $U=-2.5\,10^{-4}$. }
	\label{fig:Polarization_pattern}
\end{center}
\end{figure}

Out of the $\Gamma$ point, the electric field $\vec{E_\pm}$ in Cartesian basis can be written as:
\begin{equation}
    \vec{E_\pm} = E_{x,\pm}\vec{u_x} + E_{y,\pm}\vec{u_y}  
\end{equation}
Here $E_{x,\pm}$ and $E_{y,\pm}$ are complex amplitudes, given by:
\begin{subequations}
\begin{align}
    E_{x,\pm}&=E_0\left(\cos\theta_{+1} + C_\pm \cos\theta_{-1}\right),\label{eq:Ex}\\
    E_{y,\pm}&=E_0\left(\sin\theta_{+1} + C_\pm \sin\theta_{-1}\right).\label{eq:Ey}
\end{align}
\end{subequations}
 The orientation angle $\phi_\pm$ of $\vec{E_{\pm}}(q_x,q_y)$ with respect to $\vec{u_x}$, and the ellipticity angle $\chi_\pm$ (see Fig.\ref{fig:polarization_angle}a) are given by:
\begin{subequations}
\begin{align}
    \tan2\phi_\pm&=\frac{2\text{Re}(E_{x,\pm}^*.E_{y,\pm})}{|E_{x,\pm}|^2-|E_{y,\pm}|^2}, \label{eq:phi}\\
    \sin2\chi_\pm&=\frac{2\text{Im}(E_{x,\pm}^*.E_{y,\pm})}{|E_{x,\pm}|^2+|E_{y,\pm}|^2}.     \label{eq:chi}
\end{align}
\end{subequations}
Numerically, we will employ \eqref{eq:Ex},\eqref{eq:Ey},\eqref{eq:phi} and \eqref{eq:chi} to calculate  the polarization texture in the momentum space. Figure \ref{fig:Polarization_pattern} depicts the polarization pattern, given by the mapping of the orientation angle $\phi_\pm$ and the ellipticity angle $\chi_\pm$, of $\ket{\pm}$. It shows clearly a polarization singularity at the $\Gamma$ point for $\ket{+}$.

The topological charge, i.e. winding number, associated to the singularity at $q_x=q_y=0$ of the vector field $\vec{E_+}(q_x,q_y)$ is defined as:
\begin{equation}
    m = \frac{1}{2\pi} \oint_\mathcal{C} d\vec{q} \nabla_{\vec{q}} \phi_+.
\end{equation}
where $\mathcal{C}$ is an arbitrary circulation encircling the singularity. Since the singularity is located at $q_x=q_y=0$ and $m$ does not depend on the choice of $\mathcal{C}$, we adopt an adequate $\mathcal{C}$ so that the integral can be easily calculated. The trick is to define the complex number $z=\rho e^{i\varphi}$ with:
\begin{subequations}
\begin{align}
 \text{Re}(z)&=\rho \cos\varphi=\frac{q_x}{2n_g\sqrt{U^2+\gamma_r^2}},\label{eq:Re_z}\\
 \text{Im}(z)&=\rho \sin\varphi=q_y. \label{eq:Im_z}  
 \end{align}
\end{subequations}
 The winding number will be calculated by encircling in the counter-clockwise direction around $q_x=q_y=0$ when fixing $\rho\ll 1$ constant and varying $\varphi$. The condition $\rho\ll 1$ implies that $\lvert q_x \rvert,\lvert q_y \rvert\ll 1$. Thus from \eqref{eq:theta},\eqref{eq:C}, we can approximate:
\begin{equation}
    \begin{cases}
    \sin\theta_{+1}\approx 1 \\
    \cos\theta_{+1}\approx -\rho \sin\varphi \\
     C_+\approx  1 - 2\rho \cos\varphi\frac{U+i\gamma_r}{\sqrt{U^2+\gamma_r^2}}
    \end{cases}
\end{equation}
Using \eqref{eq:Ex},\eqref{eq:Ey}, we thus have:
\begin{align*}
E_{x,+}&\approx -2\rho\sin\varphi E_0,\\
E_{y,+}&\approx  2\rho\cos\varphi\frac{U+i\gamma_r}{\sqrt{U^2+\gamma_r^2}} E_0.
\end{align*}
Using \eqref{eq:phi}, we obtain then:
\begin{align*}
    \tan2\phi_+&=\tan2\varphi \frac{U}{\sqrt{U^2+\gamma_r^2}}\\
    &=sgn(U)\tan{2\psi},
\end{align*}
where angle $\psi$ is obtained from $\varphi$ by shrinking an ellipse of semi-major axe $\sqrt{U^2+\gamma_r^2}$ and semi-mino axe $|U|$ into a circle of radius $\sqrt{U^2+\gamma_r^2}$ (see Fig.\ref{fig:polarization_angle}b). 
Hence
\begin{equation}
    \phi_+=sgn(U)\psi.
\end{equation}
Moreover, we note that the pathway  $\mathcal{C}$ of counter-clockwise direction around $q_x=q_y=0$ corresponds to $\varphi$ and  $\psi$ both vary from 0 to 2$\pi$, due to the definition \eqref{eq:Re_z}, \eqref{eq:Im_z} and the geometrical relation between $\varphi$ and $\psi$. Thus
\begin{align}
    \oint_\mathcal{C} d\varphi &= 2\pi. sgn(U),\\  \rightarrow m&=sgn(U).
\end{align}
Interestingly, together with the eigenvalue expression from \eqref{eq:real}, this result shows that the sign of $U$ dictates both the spectral order between the BIC and leaky mode, as well as the topological charge associated to the BIC.
\begin{figure}
	\begin{center}
	\includegraphics[width=0.8 \textwidth]{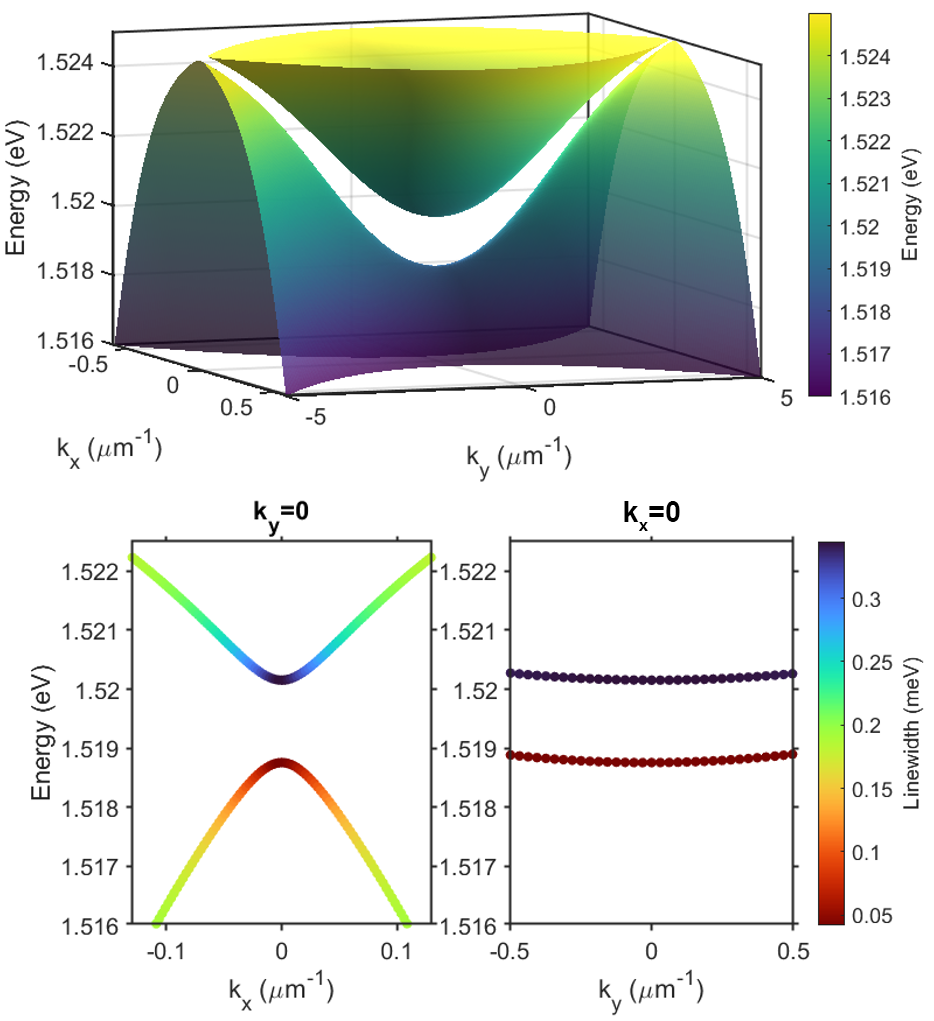}	
\caption{Dispersion characteristic of the two lower polariton modes calculated by the analytical expression \eqref{eq:LP}. Parameters: $a=243$ nm, $n_g=4$, $\hbar \omega_X=1.527$ eV, $\gamma_{nr}=2\,10^{-5}$, $\hbar \omega_0=1.526$ eV, $\gamma_r=6\,10^{-5}$, $U=-2.5\,10^{-4}$ and $V=1.4\,10^{-3}$. Here we adopt an important value of $\gamma_{nr}$ (i.e. $\gamma_{nr}=\gamma_r/3$) to highlight the quasi-BIC nature of polariton BIC. We note that to reproduce our experimental data exhibiting a lifetime of polariton BIC exceeding 300 ps, it requires a much smaller non-radiative losses corresponding to $\gamma_{nr}<\gamma_r/3000$.}
	\label{fig:dispersion_analytical}
\end{center}
\end{figure}
\subsection{Exciton-photon strong coupling regime}
We now discuss the coupling between $\ket{\pm}$ and excitonic resonances in the quantum wells. As suggested by Lu \textit{et al.}, depending on their in-plane spatial location, excitons may  efficiently interact with only $\ket{+}$ or only $\ket{-}$\cite{Lu2020}. Thus we can approximate that \ul{\textit{there are two types of excitonic resonances: one only couples to $\ket{+}$ and the other only couples to $\ket{-}$.}} Both have the same normalized frequency, given by $\hat{\omega}_X=\left(\omega_X-\omega_0\right)a/2\pi c$ and non-radiative losses, given by $\gamma_{nr}$.\\

As consequence, we can describe the strong coupling between the photonic modes and the exciton by using a model of coupled resonances according to the following Hamiltonian \cite{Lu2020}:
\begin{equation}
H_{SC}=\left(\begin{matrix}
\Omega_+ & V & 0 & 0 \\
V & \Omega_X & 0 & 0 \\
0 & 0 & \Omega_- & V \\
0 & 0 & V & \Omega_X 
\end{matrix}\right).\label{eq:H_sc}
\end{equation}

Here $\Omega_\pm =\hat{\omega}_\pm (q_x,q_y)  - i\gamma_\pm (q_x,q_y)$ is the complex energy-momentum dispersion of the two photonic modes given by \eqref{eq:Omega},\eqref{eq:real} and \eqref{eq:imag}. And $\Omega_X =\hat{\omega}_X  - i\gamma_{nr}$ is excitonic counter part. Both $\hat{\omega}_X$ and $\gamma_{nr}$  can be considered dispersionless in the range of wavevectors considered here.\\

Each $2\times2$ bloc of \eqref{eq:H_sc} gives rise to 2 polaritonic modes, one upper polariton (indicated by the lower-index $LP$),and one lower polariton (indicated by the lower-index $UP$):
\begin{align}
    \Omega_{\pm,LP}&=\frac{\Omega_\pm + \Omega_X}{2}- \sqrt{\frac{\left(\Omega_\pm - \Omega_X\right)^2}{4} + V^2},\label{eq:LP}\\
    \Omega_{\pm,UP}&=\frac{\Omega_\pm + \Omega_X}{2}+ \sqrt{\frac{\left(\Omega_\pm - \Omega_X\right)^2}{4} + V^2}.\label{eq:UP}
\end{align}
The corresponding polaritonic eigenstates are given by:
\begin{align}
    \ket{\pm,LP}&=\ket{\pm} + C_{\pm,LP}\ket{X},\label{eq:LP_eigenstate}\\
    \ket{\pm,UP}&=\ket{\pm} + C_{\pm,UP}\ket{X}\label{eq:UP_eigenstate}.
\end{align}
with
\begin{align}
    C_{\pm,LP}&=-\frac{\Omega_\pm - \Omega_X}{2V} - \sqrt{\frac{(\Omega_\pm - \Omega_X)^2}{4V^2}+1},\label{eq:A_LP}\\
     C_{\pm,UP}&=-\frac{\Omega_\pm - \Omega_X}{2V} + \sqrt{\frac{(\Omega_\pm - \Omega_X)^2}{4V^2}+1}\label{eq:A_UP}.
\end{align}
Finally, the excitonic fractions are given by:
\begin{equation}
    W_{\pm,LP(UP)}=\frac{|C_{\pm,LP(UP)}|^2}{1+|C_{\pm,LP(UP)}|^2}.
\end{equation}

Figure \ref{fig:dispersion_analytical} depicts the dispersion characteristic of the two lower polariton modes calculated by using \eqref{eq:LP}. We obtain a perfect agreement when compare to the experimental results and FDTD simulations shown in Fig.1 in the main text.\\

\subsection{Polaritonic quasi-BIC and its topological nature}
We now focus on the two polariton modes  $\ket{+,LP}$ and $\ket{+,UP}$ arise from the strong coupling between the photonic BIC $\ket{+}$ and excitons. At $q_x=q_y=0$, one may show that:
\begin{align}
    \Omega_{+,LP}(q_x=q_y=0)&=\frac{U+ \hat{\omega}_X +  i\gamma_{nr}}{2} - \sqrt{\frac{\left(U- \hat{\omega}_X -  i\gamma_{nr}\right)^2}{4} + V^2},\label{eq:LP_gamma}\\
    \Omega_{+,UP}(q_x=q_y=0)&=\frac{U+ \hat{\omega}_X +  i\gamma_{nr}}{2} + \sqrt{\frac{\left(U- \hat{\omega}_X -  i\gamma_{nr}\right)^2}{4} + V^2}.\label{eq:UP_gamma}
\end{align}
It shows that at $\Gamma$ point, the imaginary part, i.e. losses, of these modes is purely non-radiative and of excitonic origin.  As consequence, we have here quasi-BICs of quality factor only limited by the excitonic losses. Moreover, since there is no raditive losses, these polaritonic modes are \textit{dark state} which are forbidden to couple to the radiative continuum although having finite quality factor. Thus, their farfield pattern exhibits a polarization singularity at $q_x=q_y=0$. In other words, \ul{\textit{the polaritonic quasi-BICs inherit the polarization vortex in momentum space and topological nature of the photonic BIC}}.\\

Finally, one may show that the losses of polaritonic quasi-BIC at $q_x=q_y=0$ can be approximatively given by:
\begin{equation}
    \gamma_{quasi-BIC}=W_{+,LP(UP)}\times \gamma_{nr}.
\end{equation}
with $W_{+,LP(UP)}$ being the excitonic fraction. Therefore,  \ul{\textit{the quasi-BIC lifetime is limited only by the excitonic non-radiative lifetime and the excitonic fraction}}. For example, with 50$\%$ excitonic fraction, the quasi-BIC lifetime will be twice the excitonic non-radiative lifetime. The fact that we cannot obtain infinite lifetime for polaritonic quasi-BIC is the trade-off for incorporating excitonic nature such as non-linear behaviors, gain medium, interaction with electric and magnetic field into BIC physics.\\


\section{Additional experimental results}

\begin{figure}
    \centering
    \includegraphics[width=\textwidth]{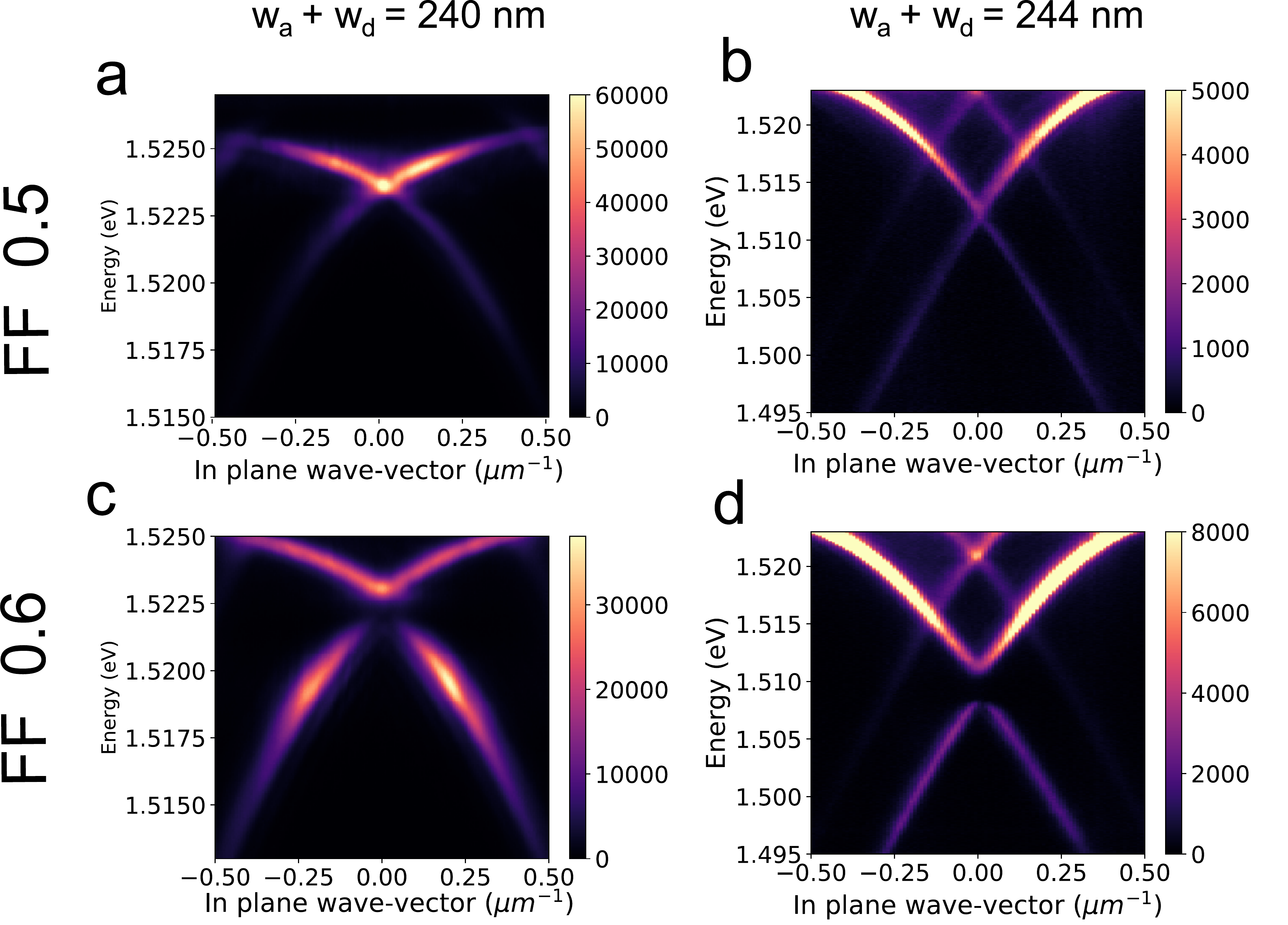}
    \caption{\textbf{(a-c):} Grating Pitch 240 nm and different FF. \textbf{(b-d)}: Pitch 244 nm and different FF;} 
    \label{Dispersion}
\end{figure}

\textbf{Effect of the gratings properties on polariton dispersions}\\
Figs.~\ref{Dispersion}(a-c) are measured on two different diffraction gratings having a fixed pitch of $w_{a} + w_{a}  = 240 nm$ but a varying fill factor (FF) of respectively $\sim 0.5$ in panel a and $\sim 0.6$ in panel c. In (a) the two counter-propagating polariton dispersions are folded by the grating and cross at $E = 1.52375$ and $k\sim 0$, no gap is present. By modifying the $FF$ of the grating, a non-zero diffractive coupling of the photonic components of the polariton modes is added, resulting in the opening of an energy gap. To show that the properties of the system are highly tunable by simply modifying the properties of the gratings, we measure polariton dispersions from another grating having a pitch of $w_{a} + w_{a} = 244 nm$  with a different $FF$ of $\sim 0.5$ and $ \sim 0.6$ respectively (Fig. \ref{Dispersion}(b-d). The different pitch modifies the folding of the dispersion resulting in a crossing point having a lower energy with respect to the crossing point of Fig.~\ref{Dispersion}(a). Again, by modifying the $FF$, an energy gap is opened with the formation of a BIC state on the lower branch.

\begin{figure}
    \centering
    \includegraphics[width=\columnwidth]{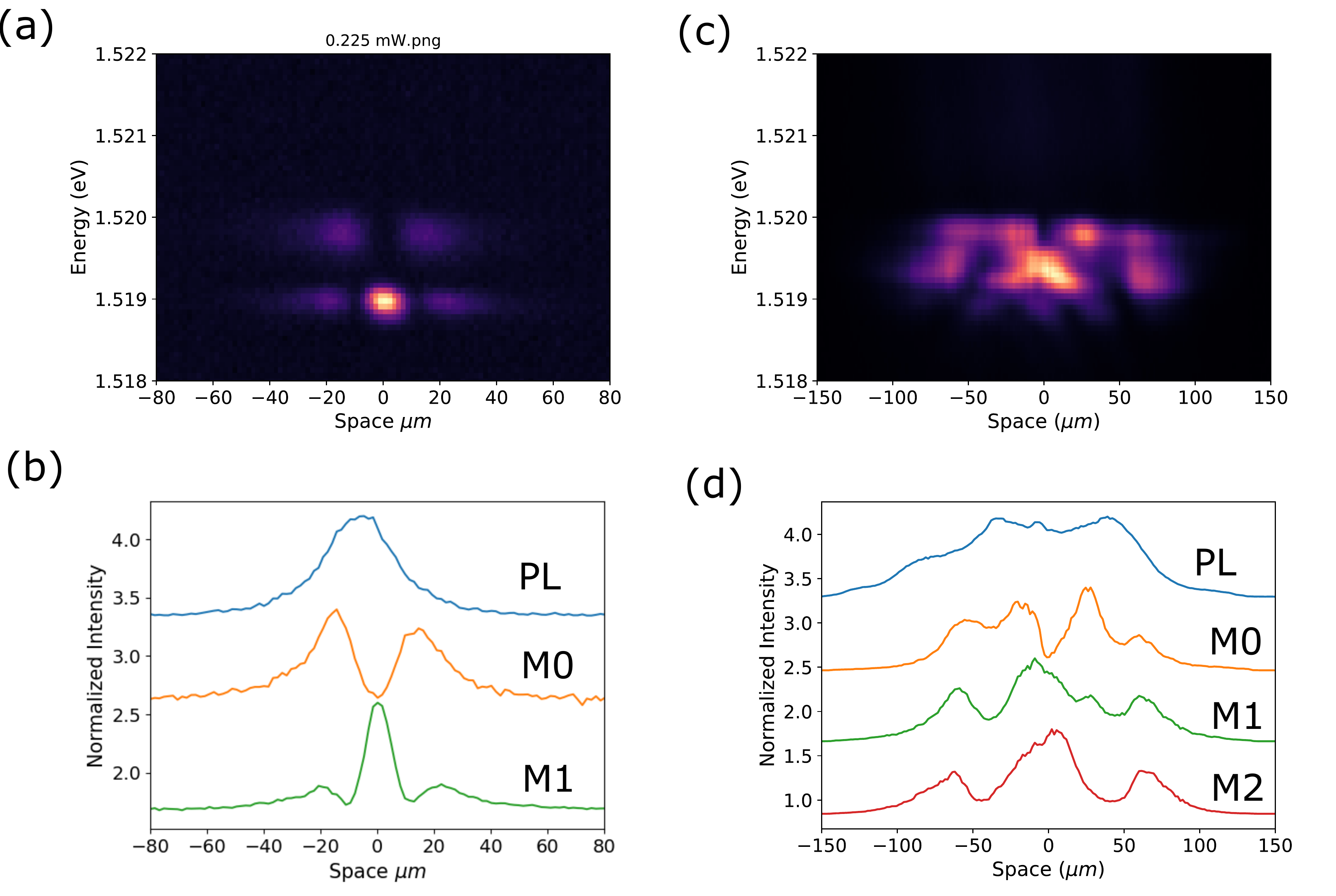}
    \caption{\textbf{(a-b):} Polariton condensation under a pumping spot of size $S \sim 30 \mu$m; for pumping densities well above the condensation threshold two modes are visible, M0 and M1 having an energy separation of about 1 meV. \textbf{(c-d)}: Polariton condensation under a pumping spot of size $S \sim 120 \mu$m; for pumping densities well above the condensation threshold three modes are visible, M0, M1 and M2, having an energy separation of about 0.3 meV.}
    \label{Spotsize}
\end{figure}

\textbf{Spatial confinement of the BIC state: effect of the spot size} \\
As described in the main text, the interacting nature of polaritons produces a blueshift of the condensate from the BIC state. The blueshift is proportional to $g_{XX}n_{X}$, with $g_{XX}$ the exciton-exciton interaction constant and $n_{x}$ the excitonic density. This means that the blueshift strongly varies in the space resulting in a local blueshift of the polariton dispersion and of the BIC emission. Due to the negative curvature of the dispersion on which the condensation from the BIC takes place, the blueshift creates an effective potential well whose size is proportional to the pumping spot size.
Figure \ref{Spotsize} (a) shows the space and energy resolved emission from polariton condensate pumped with a spot size $S = 30 \mu m$, where $S$ is extracted as the FWHM. Well above the condensation threshold, a series of condensate modes appears. The first two modes, whose profiles are represented in figure \ref{Spotsize} (b) shows an energy separation of about 1 meV. By further increasing the pumping power, other modes appear with smaller energy separation. Figure \ref{Spotsize} (c) shows the same experiment with a pump spot $S = 120 \mu m$. 
Again, above the condensation threshold a series of condensate modes appear. We note that in this case the modes are more extended in space and show a smaller energy separation, about 0.3 meV for the first three modes M0, M1 and M2. These measurements directly show that the blueshift of the polariton dispersion confines the BIC and quantizes its energy, acting as a potential well whose size is determined by the pumping spot size. 
Pumping with an extended spot highlights the global blueshift of the polariton dispersion, as all the PL is collected from under the spot in this case. This is the case of the measurements shown in Fig. 2 of the main text.

\textbf{Time-resolved measurements}
We reconstruct the temporal dynamics of the polariton condensation by using a streak camera. For each value of $k_{x}$ we acquire an energy and temporal resolved image of the emission. We scan the values of $k_{x}$ in a range comprised between $-0.3 {\mu m}^{-1}$ and $0.3 {\mu m}^{-1}$. We obtain a 3D array of ${k_{x}, Energy, Time}$ data. By plotting the data at a fixed time we obtain a snapshot of the ${k_{x}, Energy}$ emission, i.e. the polariton dispersion at a given time after the pulse arrival. 
By using this technique, we have obtained the dispersions of figure 2 (g-i) in the main text corresponding at a pumping power close to threshold. 
The videos of polariton condensation available online are obtained by plotting in sequence the snapshots corresponding to different time after the pulse arrival.

\begin{figure}
    \centering
    \includegraphics[width=\textwidth]{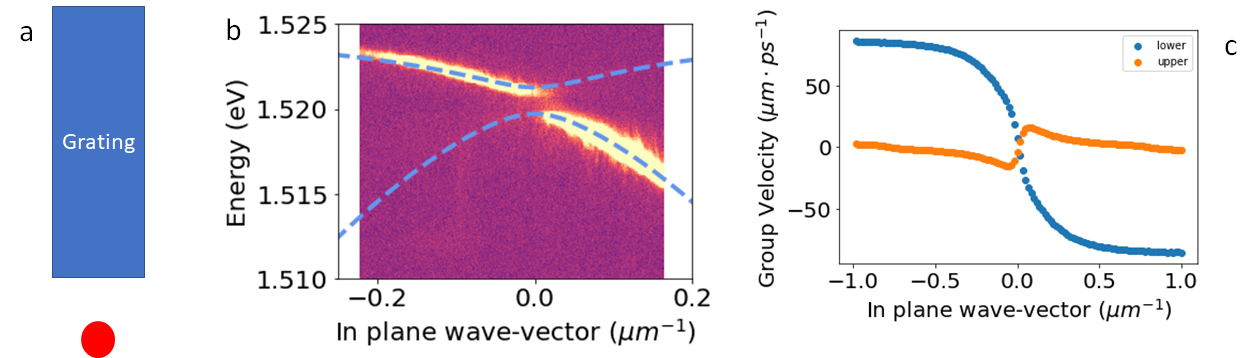}
    \caption{\textbf{a:} Excitation scheme for propagation measurements. \textbf{b}: Polariton dispersion obtained by pumping the sample with the scheme in a, only polaritons propagating upwards reach the grating. \textbf{c}: Polariton group velocity extracted from the fitted dispersion in b.} 
    \label{Prop}
\end{figure}

\textbf{Polariton Lifetime Measurements}
Given the intrinsic difficulties in quantitatively assessing polariton lifetimes by measurements of the spectral linewidths, that is inevitably limited by any inhomogeneous broadening and instrument resolution, we have extensively studied polariton propagation under non resonant injection and far away from the exciton reservoir–to avoid the effect of exciton relaxation in our system. In fact, by measuring the propagation length and by deducing the polariton propagation speed from the fitted dispersion, we are able to estimate the polariton lifetime at different energies. In particular we can measure polariton lifetimes at energies close to the BIC state, thus obtaining a lower bound for the polariton BIC lifetime (which is clearly not measurable exactly at normal incidence, where the mode is completely dark).
Polariton propagation has been experimentally studied by using the pumping scheme of figure \ref{Prop} a. In this case polaritons are created by a non-resonant pumping spot placed outside the grating. This allows the creation of a polariton population propagating upwards and extracted by the diffraction grating (the blue rectangle in figure \ref{Prop} b). The spot is placed several tens of microns away from the grating to avoid any presence of the exciton reservoir with the consequent influence on the polariton lifetime.

 The corresponding dispersion is shown in figure \ref{Prop} b. Due to the pumping scheme used, only polaritons propagating in one direction are excited. The polariton group velocity at a given in plane wave-vector can be calculated from the fitted dispersion according to the usual quantum mechanical definition, vg = (1/hbar) dE/dk. The resulting group velocity for the two polariton branches is plotted in figure \ref{Prop} c.   

\begin{figure}
    \centering
    \includegraphics[width=\textwidth]{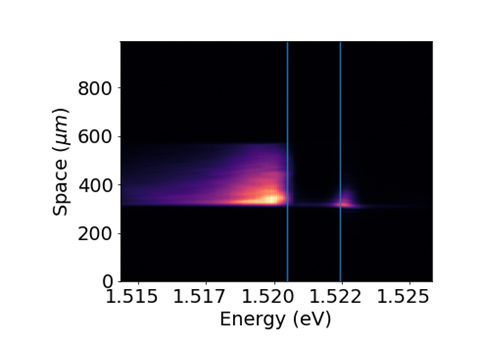}
    \caption{Energy and space resolved emission of the gratings of figure \ref{Prop}; by fitting the emission at each energy with an exponential decay we can extract the polariton decay length. Here only a portion of the bright mode is visible as it has been filetered in the reciprocal space.} 
    \label{Propreal}
\end{figure}

\begin{figure}
    \centering
    \includegraphics[width=\textwidth]{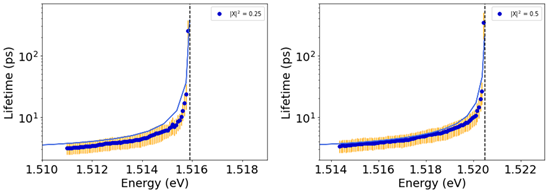}
    \caption{Polariton lifetimes for two gratings having a polariton BIC at excitonic fraction of respectively 0.25 and 0.5.} 
    \label{Polife}
\end{figure}

The spatial propagation for different polariton energies is shown in Fig. \ref{Propreal}.
As it can clearly be seen polariton propagation increases approaching the BIC state in the lower branch, while the upper branch shows very short propagation length, demonstrating the strong difference in the two polariton lifetimes across the energy gap shown by the vertical blue lines.  By fitting the spatial decay at a given energy (i.e. at a given in-plane momentum) with an exponential function, we obtain the spatial decay length at the given energy. The ratio between the decay length and the polariton group velocity of figure \ref{Prop} c gives the polariton lifetime. As a direct consequence of these measurements, we show in \ref{Polife} the resulting polariton lifetimes for two different gratings, i.e., corresponding to polariton BIC states having different excitonic fractions (0.25 and 0.5 respectively).  

In all cases, the polariton lifetimes at negligible excitonic fraction (i.e., at energies far below the BIC resonance) is of the order of few ps, corresponding to the lifetime of the bare photonic mode under the diffraction grating (i.e: of a photonic mode that is coupled to the continuum of modes outside the waveguide through the diffraction grating itself). At energy closer to the polariton BIC (vertical dashed lines), i.e. getting closer to normal incidence (or zero wave vector) in the dispersion relation, polariton lifetime strongly increases up to values of the order of hundreds of ps.    
This behavior is well reproduced by the expected lifetimes obtained from the coupled oscillator model by using the photonic modes from the FDTD (i.e.: a photonic BIC state with vanishing linewidth at k=0) and an excitonic mode having a finite linewidth corresponding to an excitonic lifetime of the order of 100 ps (see for example D. Sanvitto et al. “Rapid radiative decay of charged excitons”, PRB, 2000). The good agreement between the expected behavior modeled through coupled oscillators theory and the lifetime measurements clearly indicates that the polariton-BIC lifetime is really very long, in the order of few hundreds of picoseconds and beyond.  
We would also like to stress that the points that are closest to the BIC are quite noisy due to both the lack of signal and the error on the estimation of the group velocity when this is very close to zero, i.e., in the proximity to the dark state (see Fig. \ref{Prop}c).  We note that a significant dependence of the polariton lifetime with the excitonic fraction is not found. This may be caused by the difficulties to reach the real lifetime of the BIC state at k=0 as well as the long exciton lifetimes in wide GaAs QWs (i.e. the low nonradiative losses). 

\begin{figure}
    \centering
    \includegraphics[width=\textwidth]{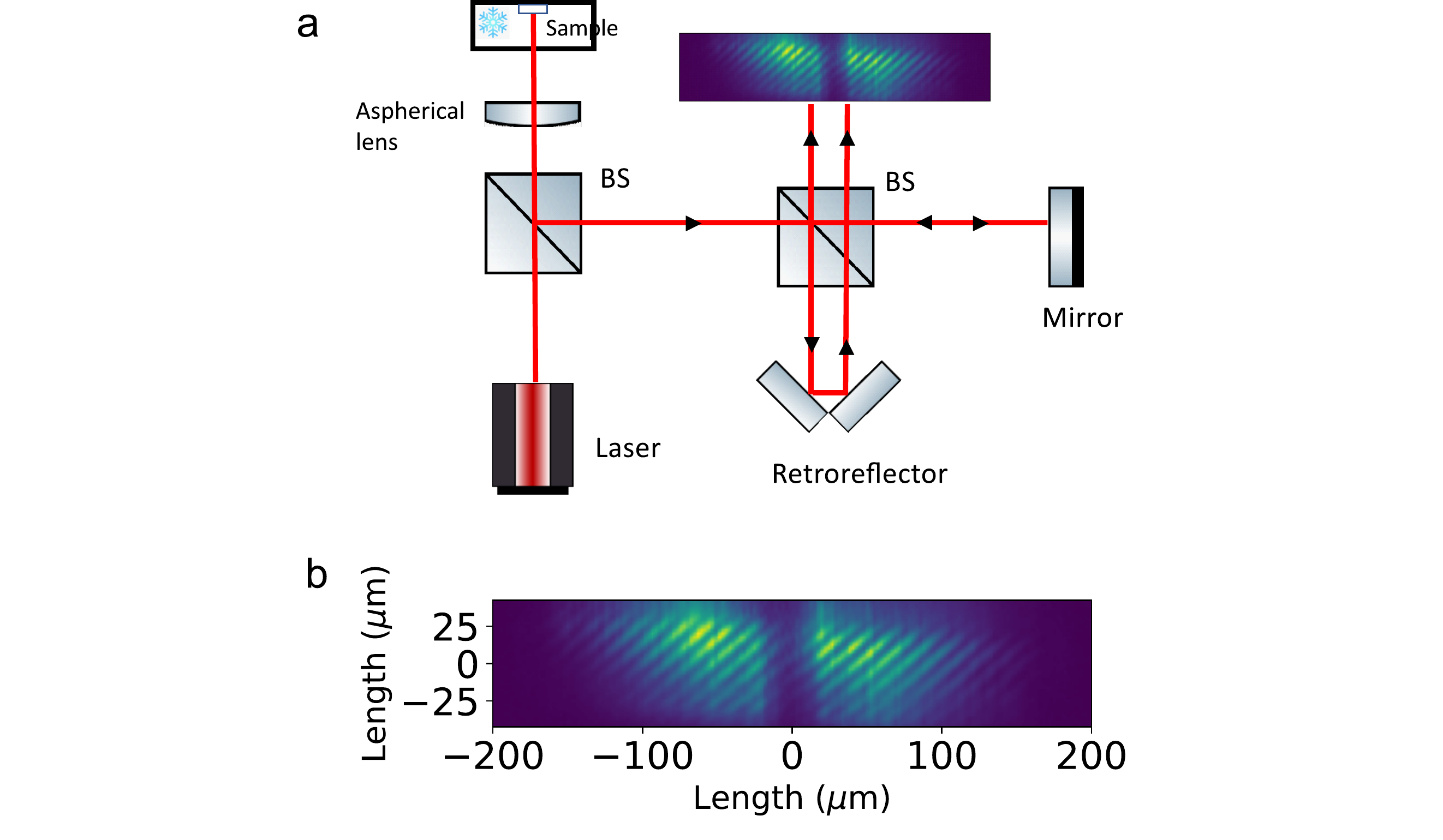}
    \caption{\textbf{a:} Scheme of the experimental setup used to measure the interferogram shown in \textbf{b}.} 
    \label{g1}
\end{figure}

\textbf{Condensate spatial coherence}
To characterize the spatial coherence properties of the condensate we have measured the interferogram from the polariton condensate through a Michelson interferometer in which one of the mirrors has been replaced by a retroreflector in order to inverse one of the images as it is shown in Fig. \ref{g1} a, see also Nature 443, 409, 2006. The two beams leaving the Michelson interferometer are then recombined at the CCD, producing the interferogram shown in figure \ref{g1} b. 
The interference fringes are visible over the entire length of the diffraction grating (300 µm in this case) demonstrating full spatial coherence. The fringe visibility is measured by the first order correlation function of Fig. 3 in the main text.   

\begin{figure}
    \centering
    \includegraphics[width=\columnwidth]{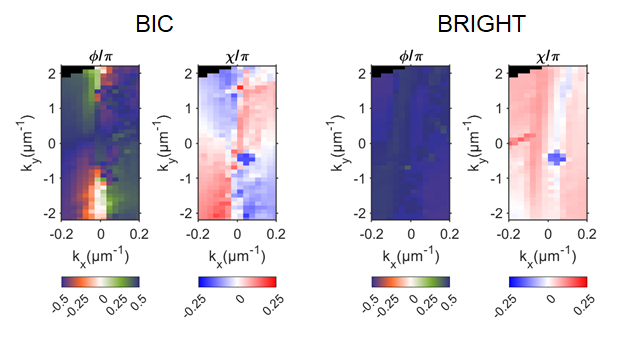}
    \caption{\textbf{Left column:} Measured polarization parameters from the BIC (first row) and the bright state (second row), $\phi$ is the polarization axis direction and $\chi$ is the ellipticity.}
    \label{Polar}
\end{figure}

\textbf{Polarization resolved measurements} \\
Figure \ref{Polar} shows the polarization parameters obtained by measuring the six polarization components: vertical, horizontal, diagonal, anti-diagonal, circular right and circular left. From these six polarization components we compute the Stokes parameters and hence the polarization ellipse direction $\phi$ and the ellipticity $\chi$ shown in the column "Experiment". We note that the polarization vortex is visible only on the BIC state while the bright state is essentially polarized along the TE polarization (black in the colorscale used in the images).  

\end{document}